\documentclass[preprint,authoryear,3p,times]{elsarticle}

\usepackage[latin1]{inputenc}

\usepackage{graphicx}
\usepackage{subfigure}

\usepackage{bm}
\usepackage{hyperref}

\usepackage{empheq}
\usepackage{xspace}
\usepackage{color}
\usepackage{amssymb}
\usepackage{booktabs}
\usepackage{appendix}

\hypersetup{
bookmarksnumbered = true,
unicode=false,          
pdftoolbar=true,        
pdfmenubar=true,        
pdffitwindow=false,     
pdfstartview={Fit},    
pdftitle={Motility initiation in active gels},    
pdfauthor={P.Recho},     
pdfsubject={Subject},   
pdfcreator={Creator},   
pdfproducer={Producer}, 
pdfkeywords={keyword1} {key2} {key3}, 
pdfnewwindow=true,      
colorlinks=true,       
linkcolor=black,          
citecolor=black,        
filecolor=black,      
urlcolor=black,           
pdfdisplaydoctitle = true
}

\renewcommand{\bar}{\overline}

\newcommand{\Z}{\mathcal{Z}}
\newcommand{\K}{\mathcal{K}}
\renewcommand{\P}{\mathcal{P}}

\usepackage[normalem]{ulem}
\usepackage{color}
\definecolor{myorange}{rgb}{0.9568,0.4941,0.1961}
\definecolor{myred}{rgb}{0.9098,0.1294,0.2078}
\definecolor{myblue}{rgb}{0.0352,0.4981,0.6509}
\definecolor{mygreen}{rgb}{0.2235,0.6353,0.2588}

\usepackage{marvosym}

\journal{Journal of the Mechanics and Physics of Solids}

\usepackage[normalem]{ulem} 

\begin{document} 
\text

\title{Mechanics of motility initiation and motility arrest in crawling cells }

\author[lms,curie]{P. Recho}
\author[bristol]{T. Putelat}
\author[lms]{L. Truskinovsky}
\ead{trusk@lms.polytechnique.fr}
\address[lms]{
LMS,  CNRS-UMR  7649,
Ecole Polytechnique, Route de Saclay, 91128 Palaiseau,  France}
\address[curie]{Physicochimie Curie, CNRS-UMR168, Institut Curie, Centre de Recherche, 26 rue d'Ulm F-75248 Paris Cedex 05, France}
\address[bristol]{
DEM, Queen's School of Engineering, University of Bristol, Bristol, BS8 1TR, UK}

\begin{abstract}
Motility initiation in crawling cells requires transformation of a symmetric state into a polarized  state. In contrast,  motility arrest is associated with re-symmetrization of the internal configuration of a cell.  Experiments  on keratocytes suggest that polarization is  triggered  by  the increased contractility of motor proteins but the conditions of  re-symmetrization  remain unknown.  In this paper we  show that if adhesion with the extra-cellular substrate is sufficiently low, the progressive intensification of motor-induced  contraction  may be  responsible  for  both  transitions: from static (symmetric) to motile (polarized)  at a lower contractility threshold and from motile (polarized) back to static (symmetric) at a higher contractility threshold.  Our model of lamellipodial cell motility  is based on a  1D projection of the complex intra-cellular dynamics on the direction of locomotion. In the interest of analytical transparency we also neglect  active protrusion and view adhesion as passive. Despite  the unavoidable oversimplifications
 associated with these assumptions, the model reproduces  quantitatively the motility initiation pattern in fish keratocytes and 
 reveals  a crucial role
played in cell motility by the  nonlocal  feedback  between the 
 mechanics  and the transport of active agents.   A  prediction of the model that a crawling cell can stop and re-symmetrize when contractility  increases sufficiently far  beyond the motility initiation threshold  still awaits experimental verification. 
\end{abstract}

\maketitle

\section{Introduction}

The ability of cells to self-propel  is essential for many biological processes.     In the early life of an embryo, stem cells  move to form tissues and organs. During the immune response, leukocytes  migrate through capillaries to attack infections. Wound healing requires the motion of epithelial cells.  While  the biochemistry of such motility is rather well understood,  the underlying \emph{mechanics} of  active continuum media  is  still in the stage of development \citep{Bray2000, Mogilner2009, Carlsson2008, Joanny2011, Adler2013, Ziebert2013, Giomi2014, Recho2014, Cox2014}. 

At a very general level, a cell can be viewed as an elastic `bag' whose interior is separated from the exterior by a bi-layer lipid membrane. The membrane  is attached  from inside to a thin cortex -  an active muscle-type layer maintaining the cell's shape. The  interior  is filled with a passive medium, the cytosol, where all essential cell organelles are immersed. The active machinery inside the cytosol ensuring self-propulsion  is  contained in the cytoskeleton:  a perpetually renewed network of  actin  filaments   cross-linked by  myosin  motors  that can inflict contractile stresses. The cytoskeleton can be mechanically linked to the cell exterior  through  adhesion proteins  \citep{Alberts2002}. 

The elementary mechanisms responsible for the steady crawling of keratocytes (flattened cells with fibroblastic functions)  have been identified \citep{Abercrombie1980, Bell1984, Stossel1993, Bellairs2000}. The advance starts with protrusion driven by active polymerization of the actin network in the frontal area of the cell (the lamellipodium)  with  a simultaneous formation of adhesion clusters at the advancing edge. After the adhesion of the protruding part of the cell is secured, the cytoskeleton contracts thanks to the activity of myosin motors. This contraction leads to the detachment at the rear  and lessening of the actin network through de-polymerization.  All these \emph{active} phenomena are driven by ATP hydrolysis and are highly synchronized which allows the cell to move  with a stable shape and relatively constant velocity \citep{Barnhart2011}. 

The \emph{initiation} of  such motility requires a polarization of the cell  which is a process that discriminates the leading edge from the trailing edge. The implied symmetry breaking turns  a  symmetric stationary configuration of a cell  into a polar motile  configuration. While both contraction and protrusion contribute to steady state cell migration, contraction appears to be the dominating mechanism of polarization: it has been shown experimentally that motility initiation in keratocytes may be triggered by raising the contractility of myosin \citep{Verkhovsky1999, Csucs2007, Lombardi2007, Yam2007, Vicente-Manzanares2009, Poincloux2011}. It is also known   that cells  may  self propel   by contraction only \citep{Keller2002}.   

In physical terms, the contraction-driven polarization/motility is performed by `pullers' (contractile agents)  while  `pushers' (protrusive agents) remain largely  disabled. Some numerical models suggest that the relative role of `pushers' and `pullers'  in cellular motility may be tightly linked to  the task to be performed \citep{Simha2002, Saintillan2012}  and  even to the nature of the cargo \citep{Recho2013b}. However,  it is still not fully understood why in  case of keratocytes the   motility initiation is primarily contraction-driven. In contrast to motility  initiation, the reciprocal process of motility  \emph{arrest} is associated with re-symmetrization and such symmetry recovery is a  typical precursor of  cell division \citep{stewart2011hydrostatic, Lancaster2013,Lancaster2014}.  It is not yet clear whether re-symmetrization is also predominantly contraction-driven and if yes, whether it requires contractility reduction or  contractility increase beyond the motility initiation threshold. It is, however,  known that some cells can switch  from  static to  motile state as a result of  a decrease in the level of contractility \citep{Liu2010, Hur2011}.

A large variety of modeling approaches targeting  cell motility  mechanisms can be found in the  literature, see the reviews by \cite{Rafelski2004, Carlsson2008, Mogilner2009, Wang2012}.  In some models, the actin network  is viewed as a highly viscous active \emph{fluid} moving through a cytoplasm  by generating internal contractile stresses \citep{Alt1999, Oliver2005, Herant2010, Kimpton2014}. In other models, the cytoskeleton is represented by an active \emph{gel} whose polar nature is modeled  in the framework  of  the theory of  liquid crystals \citep{Kruse2005, Joanny2007, Julicher2007, Joanny2011, Callan-Jones2011}.  The active gel theory approach, which we basically follow in this study without an explicit reference to local orientational order,  was particularly successful in reproducing rings, asters, vortices and some other  sub cellular structures observed in vivo \citep{Doubrovinski2007, Sankararaman2009, Doubrovinski2010, Du2012}. At sufficiently fast time scales, the cytoskeleton can be also modeled as  an active \emph{solid} with a highly nonlinear  scale-free rheology \citep{Broedersz2014,Pritchard2014}. 

Various  specific sub-elements of the motility mechanism have been   subjected to careful mechanical study.  Thus, it was shown that in some cases the plasmic membrane with an attached cortex can be viewed as a passive elastic surface and modeled by phase field methods allowing one to go smoothly through topological transitions \citep{Wang2012,Giomi2014}.  In other cases, the membrane may  also play an active role, for instance, an asymmetric distribution of channels  on the surface of the membrane can be responsible for a particular mechanism of cell  motility  relying on  variation of  osmotic pressure \citep{Stroka2014}. While most models assume that the cell membrane interacts with the exterior of the cell through passive viscous forces,  active dynamics of adhesion complexes has recently become an area of intense research  driven in part by the   finding  of a complex dependence of the crawling velocity on the  adhesive properties of the environment \citep{Dimilla1991, Novak2004, Deshpande2008, Gao2011, Lin2008, Ronan2014, Lin2010, Ziebert2013}.  The account of other relevant factors, including  realistic geometry, G-actin transport,  Rac/Rho-regulation, etc., have  led to the development of  rather comprehensive models that can already serve as powerful predictive tools \citep{Rubinstein2009, Wolgemuth2011,Tjhung2012, Giomi2014, Barnhart2015}.

The more focused problem of finding the detailed mechanism of motility   initiation,  is most commonly addressed in the framework of theories emphasizing polymerization-driven protrusion \citep{Mogilner2002, Dawes2005, Bernheim-Groswasser2005, Schreiber2010, Campas2012, Hodge2012}. With such emphasis on `pushers', spontaneous polarization  was studied by \cite{Kozlov2007, Callan-Jones2008, John2008, Hawkins2009, Hawkins2010, Doubrovinski2011, Blanch-Mercader2013}. In \cite{Banerjee2011, Ziebert2012} and \cite{Ziebert2013}, polarization was interpreted as a result of an inhomogeneity of adhesive interactions.  Yet another group of authors have successfully argued that cell polarity may be induced  by a Turing-type instability \citep{Mori2008, Altschuler2008, vanderlei2010, Jilkine2011}.  Such a variety of modeling approaches  is a manifestation of  the fact that very different mechanisms of motility initiation are  engaged in  cells of different types.

The observation that contraction may be the leading  factor behind the  polarization of \emph{keratocytes} has been  broadly discussed in the literature. It was realized that  active contraction creates an asymmetry-amplifying positive feedback because it causes actin flow which in turn carries the regulators of contraction \citep{Kruse2003, Ahmadi2006, Salbreux2009, Recho2013, Barnhart2015}. In constrained  conditions such positive feedback  generates peaks in the concentration of stress activators (myosin motors) \citep{Bois2011, Howard2011} and  this patterning mechanism   was  used to model polarization induced by angular cortex flow  \citep{Hawkins2009a, Hawkins2011}. Closely related heuristic models of the Keller-Segel type \citep{Perthame2008} describing symmetry breaking and localization  were independently proposed by \cite{Kruse2003a, Calvez2010}.  In  all these models, however, the effect of contraction (pullers) was obscured by the account of other mechanisms, in particular,  polymerization induced protrusion (pushers), and the focus was on generation of internal flow and the resulting pattern formation,  rather than on the problem of ensuring steady \emph{translocation} of a cell. 

This shortcoming  was overcome in more recent models of contraction-induced polarization relying on splay instability in an active gel \citep{Tjhung2012, Giomi2014, Tjhung2015}. In these model, however,  `pushers' were not the only players, in particular, polarization was induced by a local phase transition from non-polar to polar gel. In \cite{Callan-Jones2013}, the motility initiation was attributed to a contraction-induced instability in a poro-elastic active gel permeated by a solvent. Here again the non-contractile active mechanism was involved as well and therefore  the domineering role of contraction could not be made explicit.

The goal of the present paper is to focus on the special role of  \emph{bare contraction} in symmetry breaking processes  by studying a  minimalistic,  analytically transparent  model of motility initiation in a segment of an active gel. Following previous work, we exploit the Keller-Segel mechanism, but now in a \emph{free boundary} setting, and show that the underlying symmetry breaking instability is fundamentally similar to an uphill diffusion of the Cahn-Hilliard type. In contrast to most previous studies,  our contraction driven \emph{translocation} of a cell is caused exclusively by the internal flow generated by molecular motors (pullers) and no other active agents are involved. Each `puller' contributes to the stress field and simultaneously undergoes biased random motion resulting in an uphill diffusion along the corresponding stress gradient. In other words our `pullers' (active cross-linkers) use the continuum environment (passive actin network) as a medium through which they interact and self-organize.  

We emphasize that the contraction mechanism of polarization and motility \citep{Recho2013,Recho2014} is conceptually very close to chemotaxis, however,  instead of \emph{chemical} gradients, the localization and motility is ultimately  driven by  the self-induced \emph{mechanical} gradients.  More specifically, the pullers  propel the passive medium by inflicting contraction which creates an autocatalytic effect since the pullers are themselves advected by this medium \citep{Mayer2010}. The inevitable  build up of mechanical gradients in these conditions is limited by  diffusion which resists the runaway and provides the negative feedback.  After the symmetry of the  static  configuration is  broken in the conditions where matter can circulate, the resultant contraction-driven flow  ensures the perpetual renewal of the network and then frictional interaction with the environment allows for the steady translocation of the cell body. 

The next  natural question is how such steady translocations can be halted. For instance, if motility initiation is contraction-driven, can motility arrest be also contraction driven and what a steadily moving cell can do in order to stop and symmetrize? Several computational models  provided an indication  that   motility initiation and motility arrest may be related to a re-entrant behavior of the same  branch of motile regimes \citep{Kruse2003a,Tjhung2012, Recho2013, Giomi2014}. To make the  link between motility initiation and motility arrest  more transparent  we study in this paper an \emph{analytically} tractable problem which captures the complexity of the underlying physical phenomena. While most of  the elements of the proposed model \citep{Recho2013,Recho2014} have been anticipated by some comprehensive computational approaches \citep[e.g.][]{Rubinstein2009}, it was previously not apparent that the initiation of motility, steady translocation and the arrest of motility can be \emph{all} captured in such a minimal setting.
 
Our model of lamellipodial cell motility is based on a 1D projection of the complex intra-cellular dynamics on the direction of locomotion. In the interest of analytical transparency, we decouple the dynamics of actin and myosin by assuming infinite compressibility of the cross-linked actin network \citep{Julicher2007, Rubinstein2009}. To ensure that the crawling cell maintains its size, we introduce a simplified cortex/osmolarity mediated quasi-elastic interaction between the front and the back of the self-propelling fragment \citep{Banerjee2012, Barnhart2010, Du2012, Loosley2012}; a comparison of such mean field elasticity with more conventional bulk elasticity models can be found in \cite{Recho2013b}. We remark that the coupling between the front and the rear of the fragment may also have an active origin  resulting from different rates of polymerization and depolymerization at the extremities of the lamellipodium \citep{Recho2013b, Etienne2014}. In other  respects  we  neglect  active protrusion (pushers). We  also view adhesion as fully passive. 
 
Despite  the unavoidable oversimplifications associated with these assumptions, we show that our model reproduces  quantitatively the motility initiation pattern in fish keratocytes and  reveals  a crucial role played in cell motility by the  \emph{nonlocal}  feedback  between the mechanics   and the transport of active agents. It also provides compelling evidence that  both, the \emph{initiation} of motility and its \emph{arrest}, may be fully controlled by the average contractility of motor proteins.  

More precisely, we show that  the increase of  contractility beyond a well defined threshold leads to a  bifurcation from a static symmetric solution of the governing system of equations (of Keller-Segel type) to an asymmetric traveling wave (TW) solution corresponding to steadily moving cells. While several TW regimes may be available at the same value of parameters,  we show that stable  TW  solutions localize motors at the trailing edge of the cell in agreement with observations \citep{Verkhovsky1999, Csucs2007, Lombardi2007, Yam2007, Vicente-Manzanares2009, Poincloux2011}. Moreover, we  show that if adhesion with the extra-cellular substrate is sufficiently low, the increase of motor-induced  contraction may  induce transition  from the steady state TW solution back  to a static solution.  This  re-symmetrization transition, leading  to the  motility arrest, can be directly associated with the behavior of keratocytes prior to cell division and our model shows that such re-entrant behavior can be ensured by `pullers' without any engagement of either active protrusion or liquid crystal elasticity.   

The paper is organized as follows. In Section~2, we present a discrete ``model of a model'' which conveys the main ideas of our approach in the simplest form. In Section~3, we develop a continuum analogue of the discrete model, study its mathematical structure and  pose the problem of finding the whole set of TW solutions incorporating both static and motile regimes. In Section~4, all static solutions of the TW problem are found analytically. In Section~5, we study the fine structure of multiple bifurcations producing motile solutions from the static ones and identify parametric regimes  when these bifurcations become re-entrant. In Section~6, we investigate numerically  the initial value problem  which allows us to qualify some of the motile TW solutions  as attractors. The reconstruction the background turnover of actin, which takes place in our model without active protrusion at the leading edge, is discussed in Section~7. In Section~8, we demonstrate that our  model can quantitatively  match the experiments carried on keratocytes. The last Section highlights our main conclusions and mentions some of the unsolved problems; three appendices contain material of technical nature.

Some of the results of this paper have been previously announced in two pre-publications \citep{Recho2013, Recho2014} but without any details. In addition to providing a necessary background for \cite{Recho2013, Recho2014}, here we develop a new discrete model, investigate the nonlocal nature of the coupling between mechanics and diffusion of active agents, give a thorough analysis of the static regimes, study the bifurcation points by using the Lyapunov-Schmidt reduction technique, investigate the non-steady problem numerically, generalize the model to account for nonlinear dependence of contractile stress on motor concentration  and provide a detailed quantitative comparison of the model with experiment.

\section{The discrete model}\label{discretemodel}

Our point of departure is a conceptual discrete model elucidating the  mechanism of contraction-driven crawling and emphasizing the role of symmetry breaking in achieving the state of steady self propulsion. This "model  of a model"   allows us to clarify the role of different components of the contraction-dominated motility machinery and  link the proposed  mechanism with the previous work on  optimization of the crawling stroke irrespective of the underlying microscopic processes \citep[e.g.][]{Desimone2012, Noselli2013}. It does not, however, address directly the main issues of motility initiation and motility arrest that require more elaborate constructions.

\begin{figure}[!h] 
\begin{center}
\includegraphics[scale=0.45]{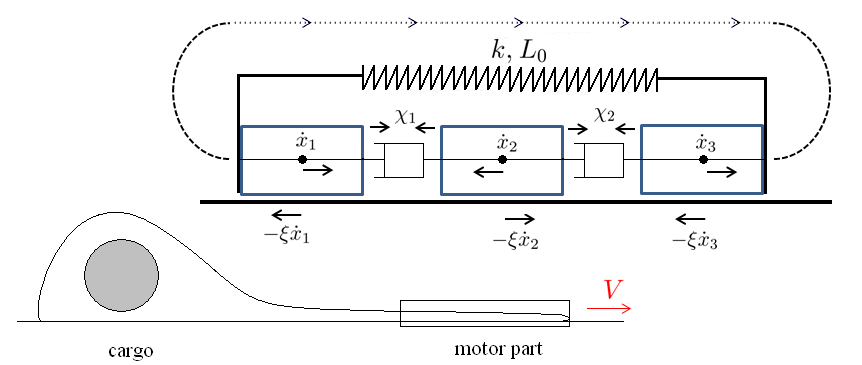}
\caption{\label{Olivia} Conceptual discrete model of the motility mechanism in a crawling keratocyte cell.}
\end{center}
\end{figure}

Recall that in crawling cells,  the `motor part' containing contracting cytoskeleton (lamellipodium),  is  a thin  active layer located close to the leading edge of the cell, see Fig.~\ref{Olivia}. We assume that all mechanical action  originates in  lamellipodium and that from the mechanical viewpoint the rest of the cell, including the nucleus, can be interpreted as cargo. The main task will be to develop a model of freely moving lamellipodium which we schematize as a segment of active gel in viscous contact with a rigid background.  The actin network  inside the gel is contracted by myosin motors which  leads to an internal flow  opposed by the viscous interaction with the background.   The unidirectional motion  in a layer adjacent to the background that ultimately propels the cell is a result of the asymmetry of contraction. 

 A toy model elucidating this point  involves three rigid blocks of size $l_b$ placed in a  frictional contact with a rigid support, characterized by the viscous drag coefficient $\xi$. The  neighboring blocks are connected by  \emph{active pullers} (force dipoles)  exerting contractile  forces. The essential long range  interactions representing  global volume constraint (due to passive elastic structures and  osmotic effects, see   Section \ref{continuummodel})  are modeled  by a linear spring  with stiffness $k$ connecting  the first and the last block. To regularize the problem we place in parallel with contractile elements additional  \emph{dashpots} characterized by the viscosity coefficient $\eta$.  In the absence of  inertia, we can then write the force balance equations in the form
\begin{equation}\label{velocity3blocks1}
 \begin{array}{c}
-l_b\xi\dot{x}_1+k\frac{x_3-x_1-L_0}{L_0}+\chi_1-\eta\frac{\dot{x}_1-\dot{x}_2}{l_b}=0\\
-l_b\xi\dot{x}_2-\chi_1+\chi_2-\eta\frac{\dot{x}_2-\dot{x}_1}{l_b}-\eta\frac{\dot{x}_2-\dot{x}_3}{l_b}=0\\
-l_b\xi\dot{x}_3-k\frac{x_3-x_1-L_0}{L_0}-\chi_2-\eta\frac{\dot{x}_3-\dot{x}_2}{l_b}=0,
\end{array}
\end{equation}
where $x_1(t), x_2(t),x_3(t)$ are the current positions of the blocks and $L_0$ is the reference length of a linear spring. This spring describes the membrane-cortex `bag' around the lamellipodium allowing the inhomogeneous contraction  to be transformed into a protruding force. We assume that polarization has already taken place and therefore the contractile force dipoles $\chi_1\geq 0$ and $\chi_2\geq 0$ acting between the two pairs of blocks are not the same  $\chi_1\neq\chi_2$. The polarization itself  requires additional constructs and will be addressed later.

System \eqref{velocity3blocks1} can be rewritten as three decoupled equations for  the length of our active segment  $L(t)=x_3(t)-x_1(t)$, its geometric center $G(t)=(x_3(t)+x_1(t))/2$ and  the position of a central block $x_2(t)$ representing the internal flow:
\begin{equation} \label{111}
 \begin{array}{c}
-l_b\xi(1+l_0^2/l_b^2)\dot{L}=\chi_1+\chi_2+2k(L/L_0-1)\\
2l_b\xi(1+3l_0^2/l_b^2)\dot{G}=\chi_1-\chi_2\\
-l_b\xi(1+3l_0^2/l_b^2)\dot{x}_2=\chi_1-\chi_2
\end{array}
\end{equation}
where $l_0=\sqrt{\eta/\xi}$ is the  hydrodynamic length scale which will ultimately play the role of a regularizing parameter.  The first equation shows that the length is converging to a steady state value:
$$L_{\infty}=L_0\left[ 1-(\chi_1+\chi_2)/(2k)\right] .$$
Notice that in order to avoid the collapse of the layer due to  contraction, it is necessary to ensure that the spring has sufficiently large stiffness
 $k>(\chi_1+\chi_2)/2.$
We also observe that independently of the  value of the evolving length $L(t)$, the  velocity of the geometrical center of our train of blocks  $V$ is always the same 
\begin{equation}\label{velocity3blocks}
V=\dot{G}=\frac{\chi_1-\chi_2}{2l_b\xi(1+3l_0^2/l_b^2)}.
\end{equation}
One can see that the system can move as a whole only if $\chi_1 \neq \chi_2$, which emphasizes  the crucial role for motility of the  \emph{polarization} and the ensuing inhomogeneity of contraction.

We observe  that the middle block moves in the  direction opposite to the motion of the center of the system with a constant velocity 
 $\dot{x}_2=-2V.$ 
\begin{figure}[!h] 
\begin{center}
\includegraphics[scale=0.5]{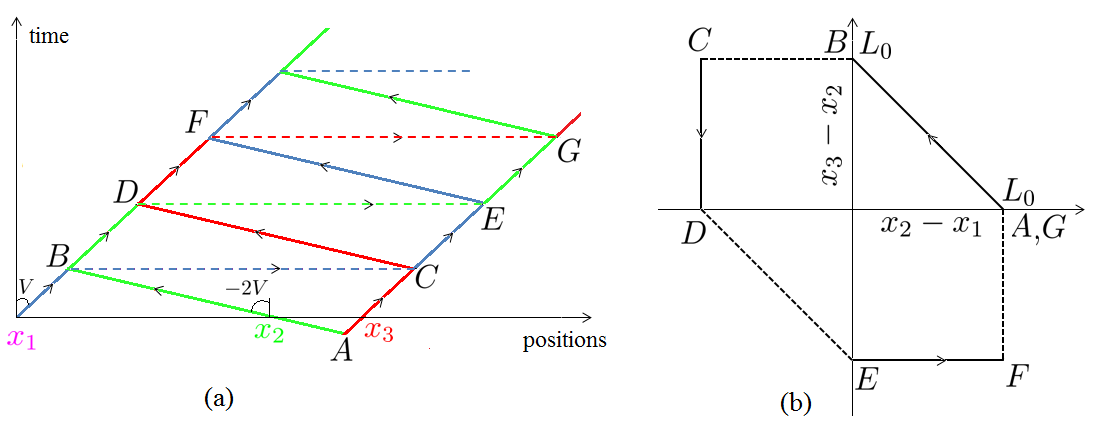}
\caption{\label{Olivia2}  (a) Schematic representation of the motion of individual particles (blocks)  forming the motor part of a crawler in a steady state regime (three particle case).  Trajectories in space time coordinates of the particles $x_1$ (magenta, $OBCEF$),  $x_2$ (green, $ABDEG$) and   $x_3$ (red, $ACDFG$); dashed lines show the jump parts of the crawling cycle. Continuous flows have to overcome friction while the jumps are assumed to be friction free. (b)  A closed loop constituting  one full stroke in the parameter space ($x_2-x_1$, $x_3-x_2$).  The time of one full stroke ($A$ to $G$) is $T_s=L_{\infty}/V$ and the distance traveled by the crawler per stroke is $VT_s=L_{\infty}$. }
\end{center}
\end{figure}
Therefore, it takes  a finite time  $\sim L_{\infty}/(3V)$ for the central block to collide with the  block at the rear. Beyond this time, system \eqref{velocity3blocks1} formally collapses and additional assumptions are needed to extend the dynamics beyond the collision point.  The origin of the problem is our focus on the layer adjacent to the rigid background and  the neglect of the global flow of actin.

To make the model more adequate we have to take into consideration that while the flow of F-actin (polymerized or  filamentous actin) is continuous  along the contact surface,  the cytoskeletal network disintegrates into  G-actin (unpolymerized monomers) at the trailing edge and reintegrates from the available  G-actin at the leading edge.  The polymerization induced depletion of G-actin at the leading edge is compensated by the diffusive counter-flow of actin monomers from the back of the cell to its front.  This counter-flow cannot be described directly in the 1D setting.

It can be modeled, however, in an indirect way by mass and momentum preserving periodic boundary conditions allowing F-actin to disappear at the rear and reappear in the front. This situation is rather typical for  continuum mechanics where unresolved spatial scales  are often replaced by balance-law-preserving jump/singularity conditions as in the case of shock waves, crack tips and boundary layers.

More specifically, to account for  global circulation (turnover)  of the cytoskeleton in a one dimensional setting,  we assume  that there is a singular source of  actin   at the front  of the cell  that is  compensated by the equivalent singular mass sink  of actin at the rear of the cell.  This assumption  allows us to close the treadmilling cycle, even though the details of the discontinuous part of the cycle, involving both the polymerization reaction and  the diffusive transport of monomers, are not explicitly resolved in the model.  We essentially postulate that there is a pool of G-actin which is replenished as fast as it is depleted and that the resulting reverse flow of actin is synchronous with the direct flow. 
  Under these assumptions the reverse flow  is viewed as passive and  and is assumed to be driven exclusively by inhomogeneous contraction.  In particular, we neglect active propulsion on free boundaries due to growth and  lessening of the network. 
  
We describe these processes in our \emph{toy model }by assuming the possibility of  creation and  destruction of the blocks.  Our goal is to account for the fact  that actin   polymerizes at the leading of the cell (where blocks are assembled) and  depolymerizes at the trailing edge of the cell (where blocks are disassembled). We offer  two interpretations of the underlying continuous  process  in terms of discrete blocks emphasizing  different sides of such passive treadmilling.  

In a \emph{first interpretation}, we assume that as a result of each collision a block at the rear is instantaneously removed from the chain and at the same time  an identical  block is  added at the front.  In other words,  each (equilibrium)  de-polymerization event at the rear is matched by an (equilibrium) polymerization event at the front.   Here we implicitly refer to   the existence of a stationary gradient of chemical potential of actin monomers  and of a large pool of monomers ready to be added to the network at the front as soon as one of them is released at the rear.  The 'instantaneous' reappearance of the disappearing blocks should be understood  as a mean to model the overall continuity of the flow.

The structure of the resulting stroke in the $t,x$ plane and in the $x_2-x_1,x_3-x_2$ plane is shown  in Fig.~\ref{Olivia2}. One can see that each block maintains its identity through the whole cycle and that its  trajectory involves a succession of continuous segments  described by  \eqref{velocity3blocks1} that are interrupted by instantaneous frictionless jumps from the rear to the front.
Notice that in this interpretation  the blocks can change order and the condition  $x_1<x_2<x_3$ is not always satisfied. For instance, when the blocks $x_1$ and $x_2$ collide at point $B$, the block $x_1$ disappears at the back (point $B$) and reappears at the front (point $C$) ahead of the block $x_3$. This jump mimics the  frictionless part of the treadmilling cycle.  Similarly, when when the block $x_3$ collides with the block $x_2$ at point $D$, the latter reappears at the point $E$ ahead of the block $x_1$.
 This interpretation is attractive because it  allows one to trace the trajectories of the blocks through subsequent treadmilling cycles. It is, however, a bit misleading because in reality the block that disappears at the back and the block which instantaneously reappears at the front are definitely not the same even though they are identical.
  
\begin{figure}[!h] 
\begin{center}
\includegraphics[scale=0.7]{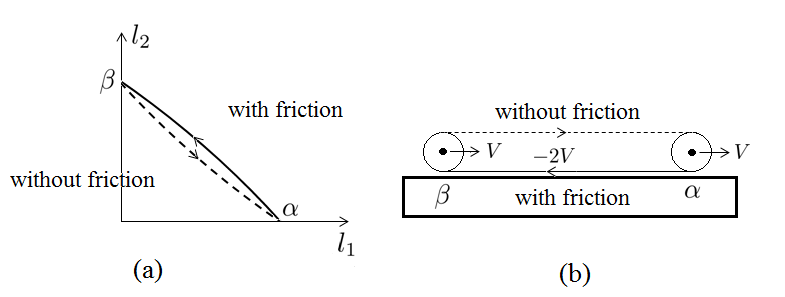}
\caption{\label{tanktread}  Schematic representation of the continuous $(\alpha,\beta)$ and the jump $(\beta,\alpha)$ part of the crawling stroke. The zero area loop in $l_1,l_2$ plane illustrating the stroke is shown in (a). The loop is not symmetric   because continuous flow have to overcome friction while the jumps are assumed to be friction free. The `tank thread'  analogy in (b) is not fully adequate because the  `departing' blocks  at  point $\beta$, that enter the pool of actin monomers, and the arriving blocks at    point $\alpha$, that are simultaneously taken from the same pool, are not the same. }
\end{center}
\end{figure}

According to a \emph{second interpretation}, illustrated in Fig.~\ref{tanktread}, the middle block is the only one to undergo cycling motion. As a result,   the ordering 
$x_1<x_2<x_3$ is  always preserved  and  the distances between the first two blocks $l_1=x_2-x_1$ and the last two blocks $l_2=x_3-x_2$  can be only positive. We can alternatively say that now  the notations  $x_1, x_2,x_3$ indicate positions only and can refer to different blocks in different times.  In this interpretation, 
  when the middle blocks hits the rear one, it is the middle block that gets recycled to the front while the rear one keeps moving continuously. 
  
In coordinates $(l_1,l_2)$ the cycle collapses on a single line, which is traveled continuously   in one direction and  discontinuously   in the other direction, see Fig.~\ref{tanktread}(a).  Notice that  the internal parameters $l_1(t)$ and  $l_2(t)$ undergo a periodic sequence of extensions and contractions which resemble the  mechanism propelling the swimming sheet \citep{Taylor1951} and  its crawling analogue \citep{Desimone2012}. The main difference is that in our case  the propulsion is achieved because of the asymmetry of friction forces acting  in the different phases of the stroke.  More specifically,  we assume that during the continuous phase of the cycle the blocks move with friction (polymerized filaments experience effective  drag transmitted by focal contacts), while during the discontinuous phase  the dissipation (associated with reaction and diffusion)  can be neglected.  The situation is remotely analogous to  that of a rotating `tank tread', see Fig.~\ref{tanktread}(b), even though in  reality the disappearing block and the appearing block are not the same.  This interpretation is closer to the physical picture where the points of the membrane (cortex) represented by two  side blocks  move with a constant speed ensuring that the cell maintains its  length.  We reiterate that both discrete interpretations are schematic  and will be backed later in the paper by an appropriate  continuum modeling.
   
Since the obtained expression for velocity (\ref{velocity3blocks})   remains finite in the limit $l_0/l_b\rightarrow 0$ it appears that the dashpots play a redundant role in this model and can be dropped.  To illustrate  the role of the dashpots we now consider the case of  $N$ coupled blocks. Then, the force balance for the central blocks $j \in [2,N-1]$ reads
$$-l_b\xi\dot{x}_j-\chi_{j-1}+\chi_{j}-\eta\frac{\dot{x}_j-\dot{x}_{j-1}}{l_b}-\eta\frac{\dot{x}_j-\dot{x}_{j+1}}{l_b}=0.$$
This system of equations can be written in the matrix form,
\begin{equation}\label{matT}
\textbf{T}\dot{\textbf{x}}=\textbf{b},
\end{equation}
where we denoted by $\dot{\textbf{x}}$ the unknown vector  $\dot{x}_1,...,\dot{x}_{N}$. The tri-diagonal matrix
$$
\textbf{T}=\left[ \begin {array}{ccccc} -(2+\frac{l_b^2}{l_0^2})&1&0&0&0\\\noalign{\medskip}1&-(2+\frac{l_b^2}{l_0^2})&1&0&0\\\noalign{\medskip}0&\ddots&\ddots&\ddots&0\\\noalign{\medskip}0&0&1&-(2+\frac{l_b^2}{l_0^2})&1\\\noalign{\medskip}0&0&0&1&-(2+\frac{l_b^2}{l_0^2})\end {array} \right]
$$
describes the viscous coupling and  frictional interaction with the background while the vector  
$$\textbf{b}=\frac{l_b}{\xi l_0^2}\left[ \begin {array}{c} -\chi_1+\sigma_0-\frac{\xi l_0^2}{l_b}\dot{x}_1\\\chi_1-\chi_2\\\noalign{\medskip}\vdots\\\noalign{\medskip}\\\chi_{N-2}-\chi_{N-1}\\\chi_{N-1}-\sigma_0-\frac{\xi l_0^2}{l_b}\dot{x}_N\end {array} \right]$$ 
with  $\sigma_0=-k(x_N-x_1-L_0)/L_0$ carries the information about  the active forcing,  the mean field type elasticity and the boundary layer effects. To find the  solution  $\dot{\textbf{x}}$, we need to invert the matrix $\textbf{T}$ and then solve a system of two coupled linear equations $\dot{x}_1=(\textbf{R b})_1$ and $\dot{x}_N=(\textbf{R b})_N$  where  $\textbf{R}=\textbf{T}^{-1}$. The components of  the matrix $\textbf{R}$ can be found explicitly  \citep{Meurant1992} $$R_{i,j}=\frac{\cosh\left( (N+1-j-i)\lambda\right)-\cosh\left( (N+1-|j-i|)\lambda\right) }{2\sinh(\lambda)\sinh((N+1)\lambda)},$$
where $\lambda=\text{arccosh}(1+ l_b^2/(2l_0^2)).$  Knowing the `velocity field', we can now compute the  steady state value of the length  
$$L_{\infty}=L_0\left( 1-\frac{\sum_{j=1}^{N-1}\cosh(\lambda(j-N/2))\chi_j}{\sum_{j=1}^{N-1}\cosh(\lambda(j-N/2))k}\right). $$
From this formula we see again that  a finite stiffness is necessary to prevent the collapse of the system under the action of  contractile stresses: assuming for instance that $\chi_i=\bar{\chi}$  we obtain the low bound for the admissible elasticity  modulus $k>\bar{\chi}$. 

The  steady velocity $V=(\dot{x}_N+\dot{x}_1)/2$ of the geometrical center of the system  can be also computed explicitly
$$V=-\frac{l_b\sum_{j=1}^{N-1}\sinh(\lambda(j-N/2))\chi_j}{2\eta\sinh(\lambda N/2)}.$$
When $N$ is even, by denoting $M= N/2$, we can rewrite this expression in the form 
$$V=-\frac{l_b\sum_{j=1}^{M-1}\sinh(j\lambda)(\chi_{M+j}-\chi_{M-j})}{2\eta\sinh(\lambda M)}.$$ 
from where it is clear that (as in the case of three blocks) the symmetry of the vector  $\chi$ with respect to the center must be broken for the system to be able to self-propel.

If we now formally drop the dashpots by assuming that $ l_0=0$  we   obtain similar expressions for the velocity and for the steady state length as in the three block ($N=3$) case 
\begin{equation}\label{velocitylimit1}
V=\frac{\chi_{N-1}-\chi_1}{2\xi l_b}, L_{\infty}=L_0\left (1-\frac{\chi_1+\chi_{N-1}}{2k}\right).
\end{equation}
The reason behind this similarity is  that, in this limit, the `flow' fully localizes in the two boundary elements,  the only ones present in the case $N=3$. More precisely,  the solution of the discrete problem depends singularly on the ratio $l_b^2/l_0^2$ and becomes progressively more concentrated around the boundary elements as  $ l_b^2/l_0^2 \rightarrow \infty$.  Such localization presents a certain analytical problem if we consider the continuum limit  when $N\rightarrow \infty$ and $l_b\rightarrow 0$ while $Nl_b\rightarrow L$, where $L$ is the continuum length of the self-propelling segment. Indeed, in this limit  the size of boundary layers tends to zero and the discrete solution converges to a distribution.  The viscosity,  introducing a  length scale $l_0$, is thus needed to preserve the regularity of  solutions in the continuum limit. 

Observe also that the limits  $ l_0 \rightarrow 0$ (dropping dashpots) and  $ l_b \rightarrow 0$ (continuum approximation) do not commute.   For instance, if we choose  in (\ref{velocitylimit1}) the motor distribution   with all $\chi_i=0$ except for one $\chi_2=\chi^{*}>0$ we obtain $V=0$ for any value of $l_b$, in particular,  when $l_b \rightarrow 0$ we still have $V\rightarrow 0$. 
If instead we first perform  the continuum limit while keeping $l_0$ finite  we obtain
\begin{equation}\label{residstresslimit}
L_{\infty}=L_0\left( 1-\frac{\int_0^{L_{\infty}}\cosh\left[ (x-L_{\infty}/2)/l_0\right]\chi(x)dx}{2kl_0\sinh\left[ L_{\infty}/(2l_0)\right] }\right) 
\end{equation}
and
\begin{equation}\label{velocitylimit}
V=-\frac{\int_0^{L_{\infty}}\sinh\left[ (x-L_{\infty}/2)/l_0\right]\chi(x)dx}{2\eta\sinh\left[ L_{\infty}/(2l_0)\right] }.
\end{equation}
If we now take a  distribution of motors $\chi(x)=\chi^{*}\delta_0$ where $\delta_0$ is the Dirac mass at $x=0$,  which can be viewed as a continuum analog of the  discrete distribution considered above,  we obtain that   $V= \chi^{*}/(2l_0^2\xi)$.
Then in the limit $l_0\rightarrow 0$ we obtain that $V\rightarrow \infty$ which is in conflict with our previous assertion that  $V=0$, obtained when the order of  limits was reversed. Assume now that $l_b \sim N^{-1}$ and hence $l_b^2/l_0^2 \sim 1/(\eta N^2)$. One can see that the crossover scaling $ \eta \sim N^{-2}$ separates the two non-commuting  limiting regimes.  For  $ l_b^2/l_0^2 \rightarrow \infty$ (which is a dimensionless version of  $ \eta \ll N^{-2}$)  the internal  flow localizes in the boundary layers whose thickness disappears when $\eta\rightarrow 0$; when we dropped the dashpots in the three element model we could not detect this localization  because  the two boundary links were the only ones present in the system.  In the other limit $ l_b^2/ l_0^2 \rightarrow 0$ (dimensionless version of $ \eta \gg  N^{-2}$)  the viscosity dominates the dynamics and the internal flow becomes uniform.

In the next sections the formulas \eqref{residstresslimit} and \eqref{velocitylimit} will be obtained directly from the continuum  model. We will also see more clearly how the introduction of the viscosity-related internal length scale and the associated nonlocality regularizes the continuum model which otherwise has only singular solutions.

\section{The continuum model}\label{continuummodel}

\begin{figure}
\begin{center}
\includegraphics[scale=0.5]{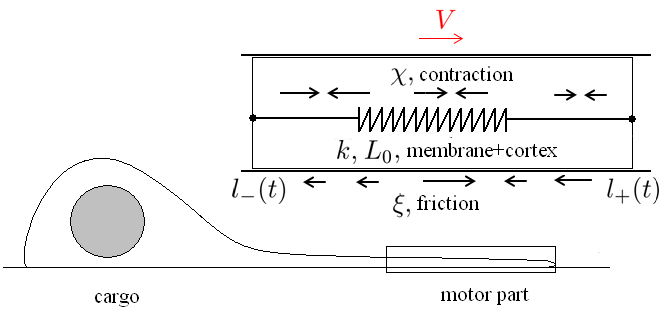}
\caption{\label{schema}  Schematic representation of a  continuum model simulating lamellopodial contraction-driven crawling.}
\end{center}
\end{figure}

We model  the lamellipodium as a one dimensional continuum layer in frictional contact with a rigid background, see Fig.~\ref{schema}.  Assuming that the drag is viscous and neglecting inertia we can write the force balance  in the form 
\begin{equation}\label{stresseq}
\partial_{x}\sigma =\xi v,
\end{equation}
where $\sigma (x,t)$ is the axial stress  and $v (x,t)$ is the velocity of the cytoskeleton (actin network). Eq. \eqref{stresseq} is the continuous analog of the  system \eqref{matT} in the discrete problem.

As in the discrete model, we denoted  by $\xi$ the coefficient of viscous drag.  Such representation  of active adhesion is  usual in the context of cell motility \citep{Rubinstein2009, Larripa2006, Julicher2007, Shao2010, Doubrovinski2011, Hawkins2011}.  It implies that the time-averaged shear stress generated by constantly engaging and disengaging focal adhesions is proportional to the velocity of the retrograde flow, see \cite{Tawada1991} for a microscopic justification.  There is evidence (both experimental \citep{Gardel2008, Gardel2010, Mogilner2009, Bois2011, Schwarz2012} and theoretical \citep{Dimilla1991, Mi2007}) that this assumption describes the behavior of focal adhesions accurately only when the retrograde flow is sufficiently slow. The behavior of adhesion strength in the broader range of velocities is biphasic and since we  neglect this  effect, we  potentially misrepresent sufficiently fast dynamics. Observe though that  for both keratocytes and PtK1 cells the rate of lamellar actin retrograde flow varies from $5$ to $30 \, \text{nm}. \text{s}^{-1}$ in usual experimental conditions  \citep{Schwarz2012}  and in this range a direct proportionality relationship between traction stress and actin retrograde flow has been confirmed experimentally \citep{Gardel2008, Fournier2010, Barnhart2011}.  The characteristic velocity of the flow in our problem is $ 20 \, \text{nm}. \text{s}^{-1}$ which falls well into the aforementioned interval where  the biphasic behavior can be neglected.

Following \cite{Kruse2006, Julicher2007, Bois2011} and \cite{Howard2011}, we assume that the cytoskeleton is a viscous gel with active pre-stress. We neglect the bulk elastic stresses that relax over a time scale of $1-10 \, \text{s}$  \citep{Rubinstein2009, Wottawah2005, Kole2005, Panorchan2006, Mofrad2009, Recho2013b} which is much shorter than characteristic time scale of motility experiments (hours). We can then describe the constitutive behavior of the gel in the form 
\begin{equation}\label{loicomportement}
\sigma=\eta \partial_{x}v+\chi c,
\end{equation}
where $\eta$ is the bulk viscosity, $c(x,t)$ is the mass concentration of motors and  $\chi>0$ is a contractile pre-stress (per motor) representing internal activity. The constitutive relation (\ref{loicomportement}) generalizes the parallel bundling of dashpots and contractile units in the discrete model.  The important new element is that   the strength of the contractile elements may now vary in both space and time. 

In the discrete model the concentration of motors $c$ was a given as a function of $x$. To obtain a more self consistent description we assume that the function $c(x,t)$ satisfies a convection-diffusion equation \citep{Rubinstein2009, Bois2011, Barnhart2011, Wolgemuth2011, Hawkins2011} 
\begin{equation}\label{drift-diffusion}
\partial_t c+\partial_x(c v)=D\partial_{xx}c,
\end{equation}
where $D$ is the diffusion coefficient. Behind (\ref{drift-diffusion}) is the  assumption that  myosin motors, actively cross-linking the actin network, are advected by the network flow and  can also diffuse which accounts for thermal fluctuations.  

To justify this  model,  consider a simple mixture model with two species representing attached and detached motors. The attached motors are  advecting  with the velocity of actin filaments and can  detach. The detached motors are freely diffusing, and  can also attach.
Suppose that the attachment-detachment process can be described by a first order kinetic equation. Then the system of equations governing the evolution of the concentrations of attached $c_a$ and detached $c_d$ motors can be written as:
$$
\begin{array}{c}
\partial_tc_a+\partial_x(c_av)=k_{\text{on}}c_d-k_{\text{off}}c_a\\
\partial_tc_d-\tilde{D}\partial_{xx}c_d=k_{\text{off}}c_a-k_{\text{on}}c_d\\
\end{array}
$$
where  $k_{\text{on}}$ and $k_{\text{off}}$ are the chemical rates  of attachment and detachment and $\tilde{D}$ is the diffusion coefficient of detached motors in the cytosol. 
Now suppose that the attachment-detachment process is chemically equilibrated  and hence 
 $ c_a/c_d=K,$  where $K=k_{\text{on}}/k_{\text{off}}$ is the reaction constant.
Then for the attached motors performing contraction we obtain
$$\frac{K+1}{K}\partial_tc_a+\partial_x(c_av)-\frac{\tilde{D}}{K}\partial_{xx}c_a=0.$$
Our equation (\ref{drift-diffusion}) is obtained   in the limit  $K\rightarrow\infty$ (fast attachment) and  $\tilde{D}/K \rightarrow  D $ (fast diffusion).  

Denote by  $l_-(t)$ and $l_+(t)$  the rear and front  boundaries of our gel segment. To account for  cortex/membrane elasticity we assume, as in the discrete model, that the boundaries are  linked through a linear spring  \citep{Barnhart2010, Du2012, Loosley2012, Recho2013b}. This assumption affects the values of the stress in the moving points  $l_-(t)$ and $l_+(t)$: 
$$\sigma(l_{\pm}(t),t) =-k(L(t)-L_0)/L_0.$$ 
Here $L(t)=l_{+}(t)-l_{-}(t)$ is the length of the  segment, $k$ is the effective elastic stiffness and $L_0$ is the reference length. As we have seen in the discrete model, the presence of  an elastic interaction plays a crucial role in preventing  the collapse of the segment due to contractile  activity of motors.

Our next assumption is that  the external boundaries of the  self propelling segment are isolated  in the sense   that they move with  the internal flow 
$\dot{l}_{\pm}=v(l_{\pm}).$ We imply here that the addition and deletion of F-actin particles inserted at the front and taken away at the rear does not contribute to propulsion. We also impose a zero exterior flux condition for  motors
 $\partial_{x}c (l_{\pm}(t),t)=0$ 
ensuring that the average concentration of motors
\begin{equation}\label{Intmot}
c_0=\frac{1}{L_0}\int_{l_-}^{l_+}c(x,t)dx
\end{equation}
is conserved. To complete the setting of the problem we need to impose the initial conditions $l_{\pm}(0)=l_{\pm}^0$ and $c(x,0)=c^0(x).$  

Our assumption that the bulk stiffness of the cytoskeleton is equal to zero (also known as the infinite compressibility assumption \citep{Julicher2007, Rubinstein2009}) allowed us to uncouple the force balance problem (which becomes statically determinate)  from the mass transport problem. Then by solving the main system of governing equations \eqref{stresseq}--\eqref{Intmot} we can obtain  the velocity field and the concentration of motors.   To recover the mass distribution of the cytoskeleton we need to solve  a decoupled  mass balance equation with a kinematically prescribed velocity field \citep{Recho2013b, Recho2013}.

Indeed, suppose that by solving the system \eqref{stresseq}--\eqref{Intmot} we found the velocity field $v(x,t)$. This  means that we also know the trajectories of the free boundaries $l_-(t)$ and $l_+(t)$. To find the mass density of actin $\rho(x,t)$ in the gel, we need to solve the mass balance equation
\begin{equation}\label{massbal}
\partial_t\rho+\partial_x(\rho v)=0
\end{equation}
with initial condition $\rho(x,0)=\rho_0(x)$. Here we  neglected the diffusion of actin which is weak comparing to the diffusion of myosin. 
Now, since both  the leading and the trailing edges of the moving lamellipodium  coincide with the trajectories of  particles,  the total mass $M$ is conserved
$$
M= \int_{l_-(t)}^{l_+(t)}\rho(x,t)dx.
$$
To address the problem of continuous circulation  and to close the cycle of the cytoskeleton flow we need to interpret the points of density singularities as actin (mass) sources and sinks. In Section \ref{actinflow} we show how the solutions can be regularized if we  cut out small regularization domains around the sources and sinks and appropriately reconnect the incoming and the outgoing flows of matter.   

\paragraph*{Dimensionless problem}  If we now normalize length by $L_0$, time by $L_0^2/D$, stress by $k$, concentration by $c_0$ and density by $M/L_0$, we can rewrite the main system of equations in  dimensionless form (without changing the notations) 
\begin{equation}\label{eq.1}
\begin{array}{c}
-\mathcal{Z}\partial_{xx}\sigma+\sigma=\mathcal{P}c,\\
\partial_t c+\mathcal{K}\partial_x(c \partial_x\sigma)=\partial_{xx}c,
\end{array}
\end{equation}
Here we introduced three main  dimensionless constants of the problem:  $$\mathcal{Z}= \eta/(\xi L_0^2),$$   the ratio of viscous and elastic length scales;  $$\mathcal{K}=k/(\xi D),$$   the ratio of stiffness induced agglomeration over diffusion and  finally $$\mathcal{P}=c_0 \chi / k,$$   the dimensionless measure  of motor contractility. One can discern in (\ref{eq.1}) the structure  of the Keller-Segel system from the theory of chemotaxis \citep[e.g.][]{Perthame2008}.   The role of the  distributed chemical attractant  is played in our case  by the stress field $\sigma$  whose gradient is the driving force affecting the `colony' of myosin motors. 

The main mathematical difference between  our formulation and the standard chemotaxis problem is that we have free boundaries. Using dimensionless variables  we can  rewrite the boundary conditions in the form
\begin{align}
\dot{l}_{\pm}(t)=\K \partial_x\sigma(l_{\pm}(t),t) , \label{BCfront} \\
\sigma(l_{\pm}(t),t)=-(L(t)-1) , \label{BCstress} \\
\partial_{x}c (l_{\pm}(t),t)=0 . \label{BCcon} 
\end{align}
The integral constraint \eqref{Intmot} reduces to 
\begin{equation}\label{e:ic_dless}
\int_{l_-}^{l_+}c(x,t)dx = 1.
\end{equation}
In dimensionless variables  the mass balance equation \eqref{massbal} takes the form
$$\partial_t\rho+\K \partial_x(\rho \partial_x\sigma)=0,$$
and  the  total mass gets normalized
\begin{equation}\label{actinnorm}
\int_{l_-(t)}^{l_+(t)}\rho(x,t)dx=1.
\end{equation}

\paragraph*{Non local reformulation} Since the first of the equations $\text{(\ref{eq.1})}$ is linear, it can be solved explicitly for $\sigma$
\begin{equation}\label{stressnonlocal}
\sigma(x,t)=-\frac{(L-1)\cosh[(G-x)/\sqrt{\Z}]}{\cosh[L/(2\sqrt{\Z})]}  + \frac{\P}{\sqrt{\Z}}\int_{l_-}^{l_+}\Psi(x,y)c(y)dy,
\end{equation}
where
$$
\Psi(x,y) = \frac{\sinh[(l_+-x)/\sqrt{\Z}]\sinh[(y-l_-)/\sqrt{\Z}]}{\sinh(L/\sqrt{\Z})}-H(y-x)\sinh[(y-x)/\sqrt{\Z}].
$$
We introduced the notations:  $H(x)$  - the Heaviside function and 
$
G(t)=[l_-(t)+l_+(t)]/2 
$
is the position of the geometric center of the moving fragment. By eliminating $\sigma$ from  $\text{(\ref{eq.1})}_2$ we obtain a single  non local partial differential equation with quadratic non linearity for $c(x,t)$ 
\begin{equation}\label{nonlocc}
\begin{array}{rl}
\partial_t c(x,t)-\K(L-1)\partial_x[\theta(x)c(x,t)]
+\frac{\P\K}{\sqrt{\Z}}\partial_x(\int_{l_-}^{l_+}\varphi(x,y)c(y,t)c(x,t)dy) = \partial_{xx}c(x,t) ,
\end{array} 
\end{equation}
where  the auxiliary velocity field 
$$
\theta(x)=\frac{\sinh[(x-G)/\sqrt{\Z}]}{\cosh[L/(2\sqrt{\Z})]}
$$ 
describes  advective flow induced by the elastic coupling between the rear and the front of the active segment. The feedback  behind contraction-driven motility is  contained in the kernel
 $$\varphi(x,y)=-\frac{\cosh[(l_+-x)/\sqrt{\Z}]\sinh[(y-l_-)/\sqrt{\Z}]}{\sinh(L/\sqrt{\Z})}+H(y-x)\cosh[(y-x)/\sqrt{\Z}], 
$$
which is due to viscosity-induced bulk interactions in the system and the effect of the boundaries.  Notice that this kernel has the action/reaction symmetry 
$
\varphi(x,y)=-\varphi(l_++l_--x,l_++l_--y)
$
which is a fundamental  constraint imposed by the balance of momentum \citep{Kruse2003a, Kruse2000, Torres2010}.

\paragraph*{Inviscid limit} To distinguish the\emph{ bulk} mechanical interactions from the effects of the boundaries, 
we use the following asymptotic  expansion \citep{Ren2000} 
\begin{equation}\label{22}
\begin{array}{cl}
\varphi_b(y-x) &= \underset{L\rightarrow \infty}{\lim}\varphi(x+G,y+G)\\
&= \frac{1}{2}\left\{ \begin{array}{c}
\exp(\frac{x-y}{\sqrt{\mathcal{Z}}}) \text{ if } x-y<0\\
-\exp(\frac{y-x}{\sqrt{\mathcal{Z}}}) \text{ if } x-y>0
\end{array} \right.
\end{array} .
\end{equation}
\begin{figure}[!h]
\centering
\includegraphics[scale=0.7]{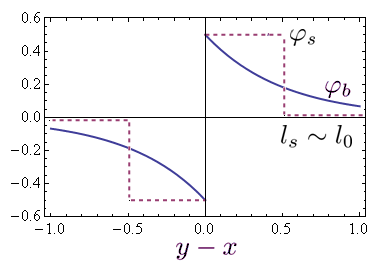}
\caption{Comparison of the bulk part of the viscosity induced interaction kernel  $\varphi_b$ (continuous line) with its mean field analog $\varphi_{s}$ (dashed line)
proposed in \cite{Kruse2003a}.
\label{compakruse} }
\end{figure}
In Fig.~\ref{compakruse}, we compare our viscosity induced interaction kernel with a long range kernel proposed in \cite{Kruse2003a, Kruse2000, Torres2010}, 
as a model of steric interactions between actin filaments with half size $l_s$ 
\begin{equation}
\varphi_s(x-y)=
\left\{\begin{array}{c}
\frac{1}{2} \text{sgn}(y-x), \text{ if } |x-y|\leq l_s\\
0, \text{ if } |x-y|>l_s\\
\end{array}\right.
\end{equation} 
The length $l_s$ plays in \cite{Kruse2003a, Kruse2000, Torres2010},  the same role as our viscous length $l_0=\sqrt{\eta/\xi}$ represented in (\ref{22}) by the dimensionless parameter  $\mathcal{Z}$.

Recall that in Section \ref{discretemodel} we  anticipated a non trivial limit in  the continuum theory when the bulk viscosity $\eta$ goes to zero. Now we see  that when $\mathcal{Z}\rightarrow 0$ the kernel $\varphi_b$  becomes singular and the nonlocality in the mechanical part of the model disappears.   From (\ref{eq.1}) we also notice that parameter $\mathcal{Z}$ enters as a coefficient in front of the highest derivative. 
Therefore, outside the boundary layers of size $\sim \sqrt{Z}$  one can formally assume that 
$\sigma=\mathcal{P}c$
which makes the main dynamic equation (\ref{nonlocc}) local
\begin{equation}\label{11}
\partial_tc(x,t)+\mathcal{PK}\partial_x(c(x,t)\partial_xc(x,t))=\partial_{xx}c(x,t).
\end{equation}
At small $\mathcal{Z}$  the non-bulk contributions to the kernel $\varphi(x,y)$ will play a role only around the extremities  of the moving segment and  in the limit $\mathcal{Z}\rightarrow 0$  will affect  only the boundary conditions.

By using  a new variable $w=1-\mathcal{KP}c$, we can rewrite   Eq. (\ref{11})  in the form 
$$
\partial_tw(x,t)+\partial_x(w\partial_xw(x,t))=0.
$$
Here we recognize the porous flow equation  which is, however, unusual because the field  $w(x,t)$ may be sign-indefinite.  In particular,  in the regimes with 
 $c>(\mathcal{KP})^{-1}$ 
one can expect an uphill diffusion similar to that of spinodal decomposition.  A systematic study of the inviscid case, requiring the   knowledge of the  boundary conditions in the limiting problem, will be done elsewhere. 
 
\paragraph*{Cell velocity} Using the boundary conditions \eqref{BCfront} we find from \eqref{stressnonlocal} an explicit formula for the (time dependent) velocity of the center of our active segment (see also Eq.~\eqref{velocitylimit} in Section~\ref{discretemodel})
\begin{equation}\label{velG}
\dot{G}=\frac{\mathcal{KP}}{2\mathcal{Z}}\int_{l_-}^{l_+}\frac{\sinh(\frac{G-x}{\sqrt{\mathcal{Z}}})}{\sinh(\frac{L}{2\sqrt{\mathcal{Z}}})}c(x,t)dx.
\end{equation}
Similarly we obtain an equation for the evolving length of the segment  (see also Eq.~\eqref{residstresslimit} in Section~\ref{discretemodel})
\begin{equation}\label{LengthL}
\dot{L}=-2\frac{\mathcal{K}}{\sqrt{\mathcal{Z}}}(L-1)\tanh(\frac{L}{2\sqrt{\mathcal{Z}}})
-\frac{\mathcal{KP}}{\mathcal{Z}}\int_{l_-}^{l_+}\frac{\cosh(\frac{G-x}{\sqrt{\mathcal{Z}}})}{\cosh(\frac{L}{2\sqrt{\mathcal{Z}}})}c(x,t)dx. 
\end{equation}
Notice that in \eqref{velG} only the odd component of the function $c(x,t)$ (with respect to the moving center $G(t)$) contributes to the integral while in \eqref{LengthL} only the even component matters. In particular,  if the concentration of motors is an even function of $x$ then $\dot{G}=0$ and the segment does not move as a whole. 
This conclusion is a direct  analog of Purcell's theorem \citep{Purcell1977} in the case of a  crawling body.  Notice that for crawling  the emphasis is made on  spatial asymmetry which replaces the emphasis on temporal asymmetry in Purcell's interpretation of swimming. 

From \eqref{velG}  we  infer that the maximal velocity of the self propelling segment  is equal to $\K\P/(2\Z)$.  If we use the data from \cite{Julicher2007, Bois2011, Howard2011}, we obtain the estimates $\chi c_0\simeq 10^3 \text{ Pa}$, $L_0=10\text{ }\mu \text{m}$ and $\eta=3\times 10^4 \text{ Pa s}$.  For the maximal velocity, this gives $\chi L_0 c_0/(2\eta)\simeq 10$ $\mu$m/min which is rather realistic in view of the data presented by \cite{Jilkine2011} and \cite{Schreiber2010}.

\paragraph*{Traveling waves}  Given our interest in  the steady modes of cell motility, which are typical for keratocytes \citep{Barnhart2011},  we need to study the  traveling wave (TW) solutions of the main system \eqref{eq.1}.  To find such solutions we assume that  the front and the rear of our segment  travel with the same 
speed $\dot{l}_{\pm}(t)\equiv V$, ensuring  the constancy of the length  $L(t)\equiv L$,  and that both the stress and the myosin concentration  depend on $x$ and $t$ through a combination $u=(x-Vt)/L$ only. Using this ansatz  we find that  the equation \eqref{eq.1}$_2$  can be solved explicitly
\begin{equation}\label{e:c_tw}
c(u)=  \frac{\exp[s(u)-VLu]}{L\int_0^1\exp[s(u)-VLu]du} .
\end{equation}
Here for convenience we introduced  a new stress variable 
$
s(u) = \mathcal{K}\left[ \sigma(u)+(L-1)\right]
$
which represents the inhomogeneous contribution  
to  internal stress field due to  active pre-stress. The  system \eqref{eq.1} reduces to the single nonlocal equation
\begin{equation} \label{eq.2}
- \frac{\Z}{L^2} s''(u)+ s(u) - \K(L-1)= \K\P \frac{\exp[s(u)-LVu]}{L\int_0^1 \exp[s(u)-VLu]du} ,
\end{equation}
supplemented by the boundary conditions
\begin{equation} \label{eq.2.bc}
s(0)=s(1)=0 \quad\text{and}\quad  s'(0)=s'(1)=LV .
\end{equation}
The two `additional' boundary conditions in \eqref{eq.2.bc} allow one to  determine  parameters $V$ and $L$ along with the function $s(u)$. After the problem (\ref{eq.2}, \ref{eq.2.bc}) is solved, the motor concentration profile can be found explicitly by using Eq. \eqref{e:c_tw}.

\section{Static regimes} \label{Sectionstatconf} 

Initiation of motility is associated with a symmetry breaking instability of a static (non-motile) configuration.  To identify  non-motile  configurations we need  to find solutions of \eqref{eq.2} with $V=0$. These solutions may still characterize the states with nontrivial active internal rearrangements of both actin and myosin. Static solutions  with periodic  boundary conditions were studied in \cite{Bois2011} and here we complement and extend their analysis.

If $V=0$, Eq. \eqref{eq.2} simplifies considerably
\begin{equation}\label{static}
-\frac{\Z}{L^2}s''+s-\mathcal{K}(L-1)=\K\P\frac{\exp(s)}{L\int_0^1\exp(s(u))du} .
\end{equation}
The nonlocal equation $(\text{\ref{static}})$ was studied extensively in many domains of science from chemotaxis \citep{Senba2000} to turbulence \citep{Caglioti1992} and gauge theory \citep{Struwe1998}. In our case,  this equation where parameter $L$ remains unknown, has to be solved with three boundary conditions $s'(0)=s(0)=s(1)=0$ because the forth  boundary condition $s'(1)=0$ is satisfied automatically.  

We begin with the study of the  regular solutions of  \eqref{static}. Instead of  $\K$ and $\P$, it will be   convenient to use another set of parameters
$A:=\K(L-1)\leq 0$ and $B:=\K\P/(L\int_0^1\exp[s(u)]du)\geq 0$.
In terms of   parameters  ($A,B$)  the problem \eqref{static}  reads   
\begin{equation}\label{staticAB}
-\frac{\Z}{L^2}s''+s-A=B\exp(s) \quad\text{with} \quad s'(0)=s(0)=s(1)=0 .
\end{equation}

\begin{figure}
\centering
\includegraphics[scale=0.3]{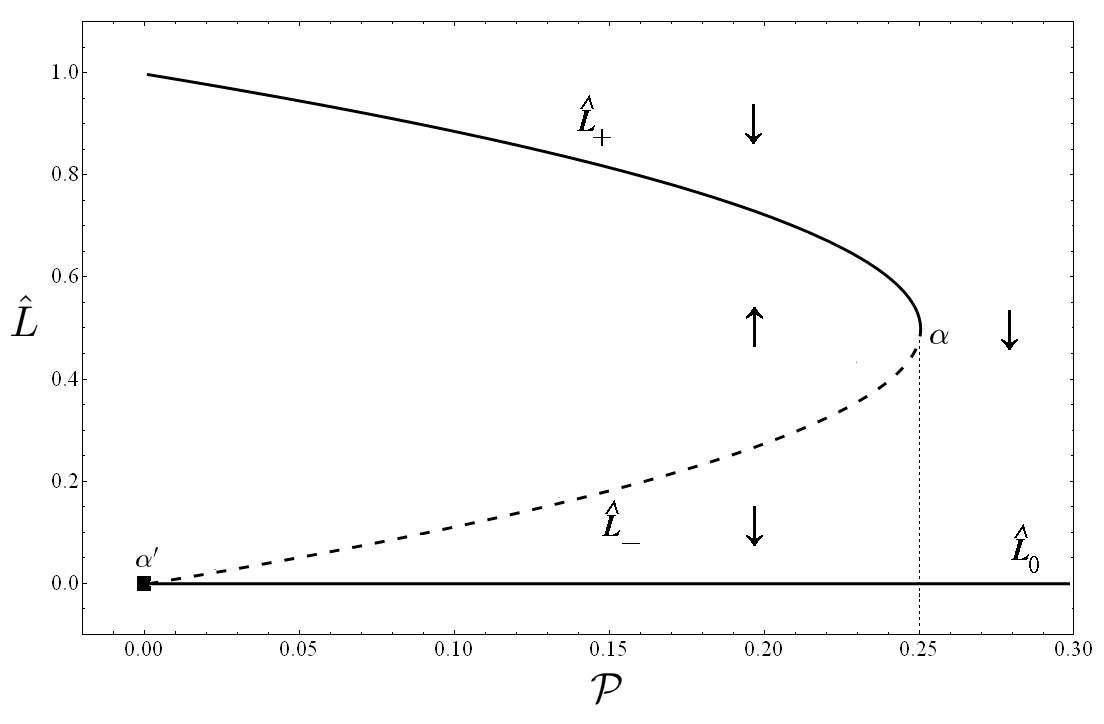}
\caption{\label{bifur2}  Three families of trivial static solutions $\hat{L}_+, \hat{L}_-$ and $\hat{L}_0$ parameterized by $\mathcal{P}$.  Solid lines show stable branches while dotted lines correspond to unstable branches. Arrows depict the basin of attraction of each branch (see Section \ref{Numericalstudy})}
\end{figure}

A trivial homogeneous solution of this problem  $s(u)=0$ exists when $A+B=0$ which is equivalent in the $(\P,\K)$ parametrization to $L=\hat{L}_{\pm}$ with,
\begin{equation}\label{nonmotileL}
\hat{L}_{\pm}=(1\pm\sqrt{1-4\P})/2.
\end{equation}
The sub-branches   with longer and shorter lengths $\hat{L}_+(\P)$ and $\hat{L}_-(\P)$, respectively, that meet at point $\alpha$ where $\hat{L}_-(\P)=\hat{L}_+(\P)$ are illustrated in Fig.~\ref{bifur2}.

\begin{figure}
\centering
\includegraphics[scale=0.5]{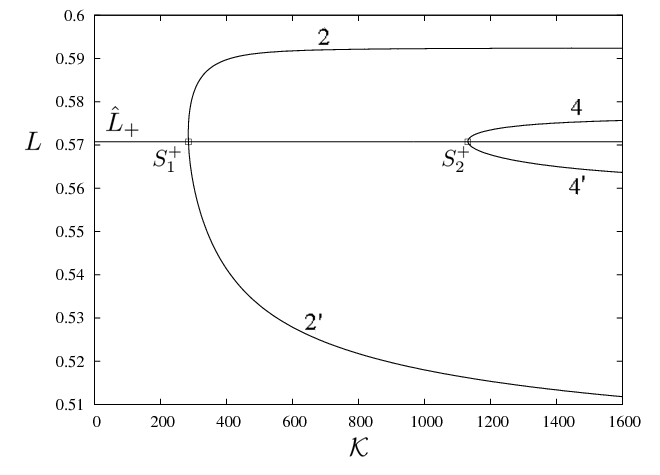}
\caption{\label{staticdiag} Bifurcation diagram for the $\hat{L}_+$ branch at fixed $\mathcal{P}=0.245$ and $\mathcal{Z}=1$. See also Fig.~\ref{bifur_K}.}
\end{figure}

To obtain nontrivial static solutions we  multiply \eqref{staticAB}  by $s'$,  integrate  and use the boundary conditions to obtain the `energy integral' 
$
s'^2= W(s) ,
$
where 
$$
W(s) =\frac{L^2}{\Z}(s^2-2As-2B\left[\exp(s)-1\right]).
$$    
The general solution of this equation can be expressed as a  quadrature,
$
u=\pm\int^{s(u)}W^{-1/2}(r)dr
$ where  we recall $u$ designates the normalized space variable. A detailed analysis of this equation is given in Appendix A, where  different families of static solutions  are  identified  as $S_{m}^{\pm}$ and $(S_{m}^{\pm})^{'}$ where the index $\pm$ specifies the $\hat{L}_{\pm}$  trivial branch from which a particular solution bifurcates: the  associated lengths  $\hat{L}_{\pm}$ are defined  in \eqref{nonmotileL}. The integer valued index $m$ corresponds to the number of spikes  in the configuration  $s(u)$.  The prime differentiates between two subfamilies belonging to the same bifurcated branch with primed subfamily having a length $L$ larger than in the trivial configuration and non-primed subfamily having the length $L$ smaller than in the trivial configuration. Fig.~\ref{Conportrait} illustrates the families  $S_{1}^{+}$ and $S_{2}^{+}$. For each family we plot the length of the fragment $L$ as a function of one of the controlling parameters, see Fig.~\ref{staticdiag}.
 
In addition to regular solutions described above   Eq. \eqref{static}  has  measure-valued  solutions corresponding to collapsed cells with length $\hat{L}_0=0$. 
First of all, as we see in Fig.~\ref{bifur2},  $\hat{L}_-(\mathcal{P})\rightarrow 0$ when $\mathcal{P}\rightarrow 0$ (point $\alpha'$) and  the limiting distribution of motors is concentrated on an infinitely small domain.  To characterize the asymptotic structure of the singular solutions we suppose that $L<<1$ and  that the maximum of $s$ is of order $L$. 
Then, by ignoring high order terms, we deduce from \eqref{static} a simplified boundary value problem
\begin{equation} \label{lin}
 \begin{array}{c}
-s''\approx\K\P L/(\Z\int_0^1[1+s(u)]du) \quad\text{with}\quad s'(0)=s(0)=s(1)=0.
\end{array}
\end{equation}
Then
 $
s(u)\approx\K\P L u(1-u)/(2\Z).
 $
and the remaining boundary condition $s'(0)=0$ is automatically satisfied  in the limit  $L\rightarrow 0$.  We can then conclude that the singular solutions are of the form
$$
s(x)=\lim_{L\rightarrow 0} L f(x/L)$$
where
$
f(u)=\K\P u(1-u)/(2\Z). 
$ Singular solutions of this type can be implicated in the description of cell splitting in a cortical geometry \citep{Turlier2014};  they  are also known  in  other fields where stationary states are described by equation (\ref{static}) \citep{Caglioti1992, Chen2001, Ohtsuka2002, Gladiali2012} . The presence of such  solutions is a sign that 
in a properly augmented  theory, accounting for the vanishing length, one can  expect localization with
active contraction balanced by a regularization mechanism, which may be, for instance, active treadmilling \citep{Recho2013b}. Our numerical solutions of a non-steady problem, which are naturally regularized because of the
finite mesh size (see Section~\ref{Numericalstudy}), show that  the almost singular solutions of the type described above are indeed attractors for initial data with $L<\hat{L}_-$ when $\P<1/4$.  
Moreover,  numerical experiments suggest that they are the only attractors for $\P>1/4$. This means that  even in the presence of a cortex-type spring,
an active segment  fragment  necessarily collapses after the contractility  parameter reaches  the threshold $\P_{\text{max}}=1/4$.

\section{Re-entrant bifurcations}\label{characteristicequation}

We first show that motile branches with $V\neq 0$ can bifurcate only from   trivial static solutions with $s(u)=0$, $V=0$ and $L=\hat{L}_{\pm}$. 
For $V\neq 0$ equation, multiplying \eqref{eq.2} by $s'-VL$, we find the relation:
\begin{equation}\label{secondint}
1-\exp(-LV)=LV\int_0^1\exp[s(u)-VLu] du. 
\end{equation}
Then in the  limit $V\rightarrow 0$ we obtain that $\int_0^{1}\exp(s(u))du=1$. Since static solutions $s(u)$ must be necessarily sign definite, see Appendix A, Eq. \ref{secondint} implies that the corresponding static solution can only be trivial $s(u)=0$.  As we have seen in Fig.~\ref{bifur2}, there are two non-singular families of  trivial solutions: one with longer ($\hat{L}_{+}$ family)  
and the other with shorter ($\hat{L}_{-}$ family) lengths.

\paragraph*{Bifurcation points} To find the bifurcation points along the trivial 
branch $(s=0,V=0,L=\hat{L}_{\pm}(\mathcal{P}))$, we introduce infinitesimal perturbations $\delta s(u)$, $\delta V$, $\delta L$ and  linearize  \eqref{eq.2} together with boundary conditions \eqref{eq.2.bc}.  We obtain the boundary value problem
\begin{equation}\label{bifurlineq}
\delta s''-\omega^2 \delta s=\frac{\mathcal{Z}\omega^2-\hat{L}^2}{\hat{L}^2(\hat{L}-1)}\left( \mathcal{Z}\frac{2\hat{L}-1}{\hat{L}}\omega^2  \delta L+\frac{\hat{L}^3(\hat{L}-1)}{2}(2u-1)  \delta V \right),
\end{equation}
\begin{equation}\label{e:linbcs}
\delta s(0) = \delta s(1) = 0 , \quad  \delta s'(0) = \delta s'(1) = \hat{L}  \delta V . 
\end{equation}
Here we introduced the notation  
\begin{equation}\label{e:omega_LKP}
\omega^2=(\hat{L}^2-\K\P\hat{L})/\Z. 
\end{equation}
Since $\omega=0$ at the trivial branch $\delta s=\delta V=\delta L=0$, we can assume that $\omega\neq 0$. The general solution of the problem \eqref{bifurlineq}, \eqref{e:linbcs} can be written explicitly
\begin{equation}\label{e:linsol}
\delta s(u)=C_1\sinh(-\omega u)+C_2\cosh(-\omega u)-\frac{\mathcal{Z}\omega^2-\hat{L}^2}{\omega^2\hat{L}^2(\hat{L}-1)}\left( \mathcal{Z}\frac{2\hat{L}-1}{\hat{L}}\omega^2  \delta L+\frac{\hat{L}^3(\hat{L}-1)}{2}(2u-1)  \delta V \right).
\end{equation}
From boundary conditions, non trivial solutions branch from the trivial ones if the matrix 
\begin{equation}\label{matri}
\left(
\begin{array}{cccc}
 1 & 0 & \frac{(2 \hat{L}-1) \left(\hat{L}^2-\omega^2 \mathcal{Z}\right)}{(\hat{L}-1) \hat{L}^3} & \frac{1}{2} \hat{L} \left(\frac{\hat{L}^2}{\omega^2 \mathcal{Z}}-1\right) \\
 \cosh (\omega) & \sinh (\omega) & \frac{(2 \hat{L}-1) \left(\hat{L}^2-\omega^2 \mathcal{Z}\right)}{(\hat{L}-1) \hat{L}^3} & \frac{1}{2} \hat{L} \left(1-\frac{\hat{L}^2}{\omega^2 \mathcal{Z}}\right) \\
 0 & \omega & 0 & -\frac{\hat{L}^3}{\omega^2 \mathcal{Z}} \\
 \omega \sinh (\omega) & \omega \cosh (\omega) & 0 & -\frac{\hat{L}^3}{\omega^2 \mathcal{Z}}
\end{array}
\right) ,
\end{equation}
has a zero determinant.  This gives a transcendental equation for  $\omega$ 
\begin{equation}\label{33}
2\hat{L}[\cosh(\omega)-1]-\K\P\omega\sinh(\omega)=0.
\end{equation}

The detailed analysis of this equation is presented  in Appendix B. 
\begin{figure}
\centering
\includegraphics[scale=0.45]{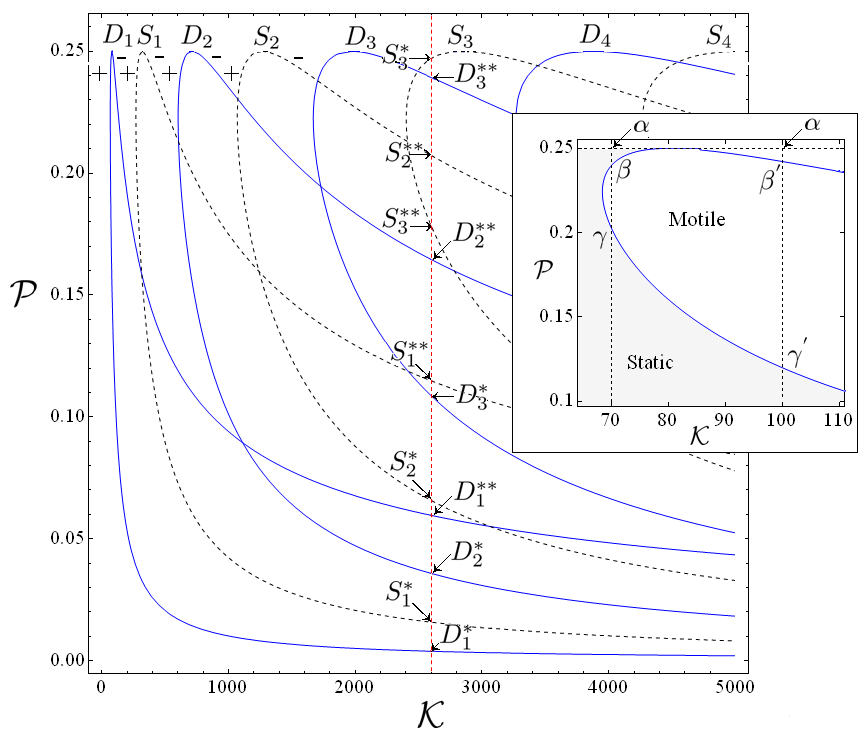}
\caption{\label{bifur1} Locus of the bifurcation points in the $(\mathcal{K},\mathcal{P})$ plane for $\mathcal{Z}=1$. Insert shows a zoom on the $D_1$ branch around the turning point at $\mathcal{P}=1/4$ where $\hat{L}_-$ and $\hat{L}_-$ branches meet.  The detailed bifurcation diagrams for $\mathcal{P}=0.245$ and $\mathcal{K}=70\text{ and }100$  are shown in Fig.~\ref{bifur_K} and Fig.~\ref{bifur_P} from where the meaning of labels $\beta$, $\gamma$, $\beta^{'}$ $\gamma^{'}$ becomes clear. The bifurcation points related to the cut $\mathcal{K}=2600$ (red dashed line) in the $(\mathcal{P},L)$ space are shown in Fig.~\ref{bifur2}.}
\end{figure}
The full locus of bifurcation points in the $(\mathcal{K},\mathcal{P})$ plane is  shown in Fig.~\ref{bifur1}. The lines of bifurcation points $+$ and $-$ originating on the trivial sub-branches  $\hat{L}_+$ and $\hat{L}_-$ smoothly connect at $\P=1/4$,  see Fig.~\ref{bifur2}. When parameter $\P$ is held constant while $\K$ is changing each family  $D_i$ and $S_i$ in Fig.~\ref{bifur1} is represented by  two points. For solutions  bifurcating 
from the trivial branch $\hat{L}_{+}$, we have $\K_+=(\hat{L}_+^2-\Z\omega^2)/(\P\hat{L}_+)$,
which gives points  $D_1^{+},S_1^{+},D_2^{+},S_2^{+},\ldots$ and 
for the branch $\hat{L}_{-}$, we have $\K_-=(\hat{L}_-^2-\Z\omega^2)/(\P\hat{L}_-)$
which gives points $D_1^{-},S_1^{-},D_2^{-},S_2^{-},\ldots$  
Notice that the total number of bifurcation points increases to infinity as $\K \rightarrow\infty$. 
Now consider the  case when $\K=const$ and 
$\P$ is varied. A line $\K=const$ in the $(\mathcal{K},\mathcal{P})$ plane cuts again each line of the bifurcation points $D_i$ and $S_i$ in two points which we denote $D_1^{*},S_1^{*},\ldots$ (solutions with longer lengths) and $D_1^{**},S_1^{**},\ldots$ (solutions with shorter lengths), see Fig.~\ref{bifur2} and Fig.~\ref{bifur1}. In most cases, one of these two points is a bifurcation originating from the $\hat{L}_-$ trivial solution while the other is from the $\hat{L}_+$ trivial solution. However, as we show in the inset in Fig.~\ref{bifur1}, the two points may also bifurcate from the same branch $\hat{L}_+$. As we show later in the paper such bifurcations are of particular interest because they describe both motility initiation and motility arrest. 
\begin{figure}
\centering
\subfigure[]{\includegraphics[scale=0.3]{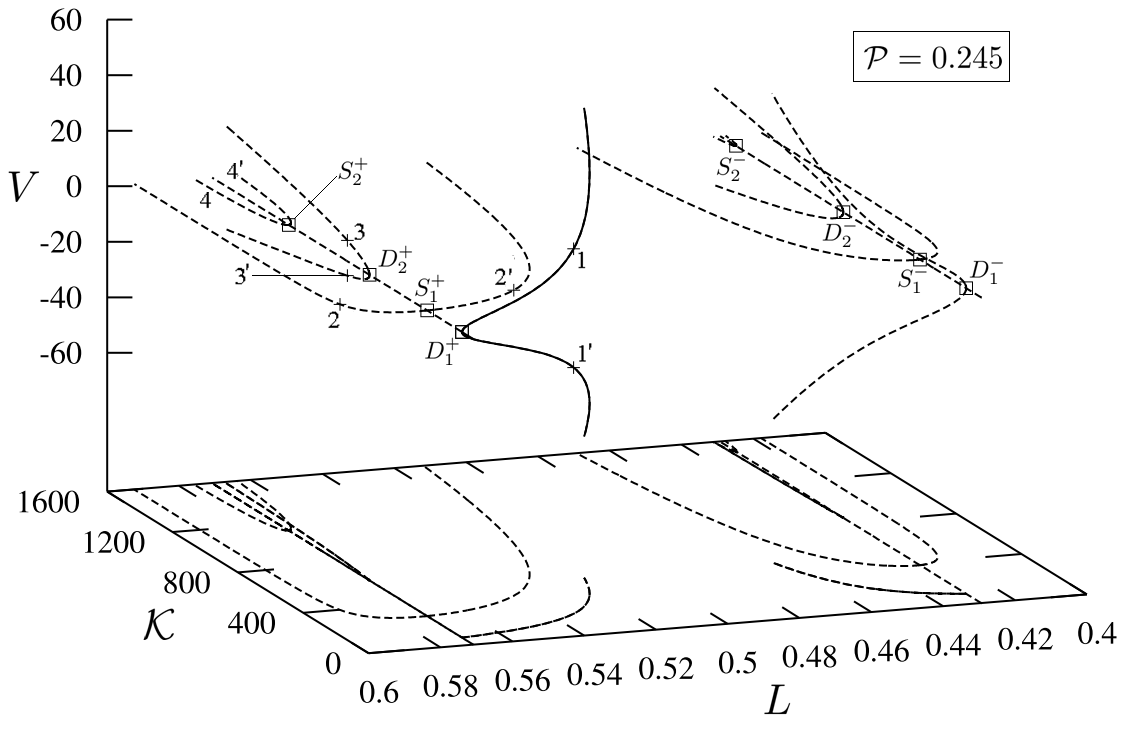}\label{bifur_K}}
\subfigure[]{\includegraphics[scale=0.55]{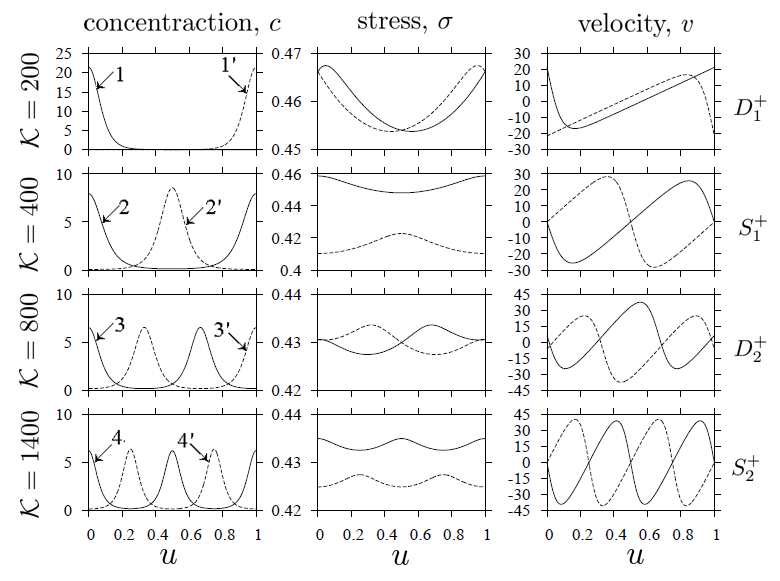}\label{profiles}}
\caption{(a) Bifurcation diagram with $\mathcal{K}$ as a parameter showing nontrivial solutions branching from families of homogeneous static solutions $\hat{L}_{+}$  and $\hat{L}_{-}$. The value $\P=0.245$ and $\Z=1$ are fixed. Solid lines show stable motile branches while all the dotted lines correspond to unstable solutions. The internal configurations corresponding to branches indicated by numbers $(1,1',2,2',\text{etc.})$ are shown in Fig.~\ref{profiles}. The projection of the bifurcation diagram on the $(\K,L)$ plane is also shown below.
(b) Internal profiles associated with successive bifurcated solutions shown in Fig.~\ref{bifur_K} for $\P=0.245$ and $\Z=1$. Our notation (1,3) correspond to asymmetric motile branches while (2,4) describe symmetric static branches.}
\end{figure}

\paragraph*{Structure of bifurcations}  After the bifurcation points are known one can use  the Lyapunov-Schmidt reduction technique to identify the nature of the corresponding bifurcations \citep{Nirenberg1974,Koiter1976, Amazigo1970}. The analysis presented in Appendix C shows that the bifurcations from the trivial to the nontrivial static branch are always transcritical. The bifurcations to motile branches can be either subcritical or supercritical. In particular, at a given  $\K$, the bifurcation from a static homogeneous solution with longer length is always supercritical  while the bifurcation from a static homogeneous 
solution with smaller length  can be either subcritical  or supercritical depending on the value of $\K$, see Appendix C.
\begin{figure}[!h]
\centering
\includegraphics[scale=0.45]{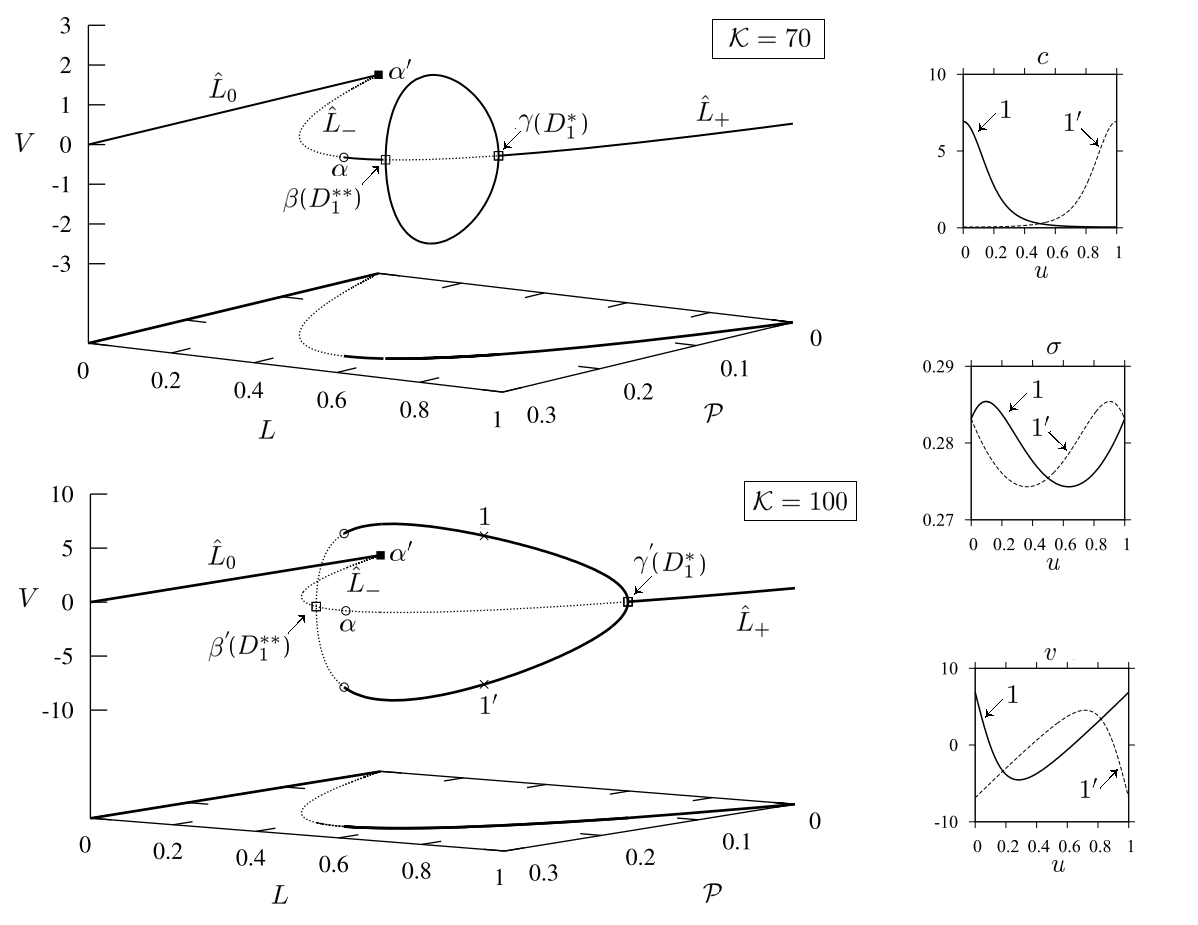}
\caption{\label{bifur_P} Bifurcation diagrams along parameter $\P$ showing motile branches connecting points $D_1^*$ and  $D_1^{**}$. 
Corresponding bifurcation points are shown in insert in Fig.~\ref{bifur1}. Solid lines show stable motile branches while all 
the dotted lines correspond to unstable solutions. The projection of the bifurcation diagram on the $(\P,L)$ plane is also shown. 
Parameter $\K$ is fixed in each graph to $\K=70$ and $\K=100$. Internal profiles on the two symmetric motile branches are also 
shown for $\K=100$. Parameter $\Z=1$.}
\end{figure}

\paragraph*{Bifurcated branches} To illustrate different types of  bifurcations we constructed the nonlinear continuation of the bifurcated branches by solving the  boundary value problem \eqref{eq.2}--\eqref{eq.2.bc}  numerically for successive values of parameters $\K$ and $\P$ (tracking algorithm, see \cite{Doedel2007}).
In Fig.~\ref{bifur_K}, we show the continuation in $\K$ for both static and motile configurations at fixed $\P$; the corresponding  profiles of motor concentration, stress and velocity are shown in Fig.~\ref{profiles}. One can see that  each  pitchfork (for motile branches) and each transcritical (for static branches) bifurcation points  gives rise to two nontrivial solutions. For instance, along the static branch $\hat{L}_+$, the bifurcation point $D_1^+$ is associated with 
two motile supercritical branches whereas the point~$S_1^+$ is associated with two transcritical static branches. 
Each pair of motile solutions is symmetric with two opposite polarizations
corresponding to two different signs of the velocity. Along the first motile branch originating at~$D_1^+$, 
the myosin  motors concentrate at the trailing edge. For the second motile branch originating at~$D_2^+$, 
there is an additional peak in the concentration profile, see Fig.~\ref{profiles}. 
In contrast, the static bifurcation point~$S_1^+$ gives rise to two symmetric configurations with different lengths 
and with myosin motors concentrated either in the middle of the cell or near the boundaries, see Fig.~\ref{profiles}. 
As one would expect, the higher order static and motile bifurcation points produce solutions with more complex internal patterns. For the branches bifurcating from the trivial configurations belonging to $\hat{L}_-$ family, the picture is similar, see Fig.~\ref{bifur_K}.

In Fig.~\ref{bifur_P}, we show in more detail the nontrivial solutions originating from the motile bifurcation points  $D_1$ at two values of parameter 
$\K$ which correspond to two sections $\alpha \beta$ and $\alpha\beta^{'}$ shown in Fig.~\ref{bifur1} (insert). 
Notice that  a single solution connects the bifurcation points $D_1^*$ (suprecritical) and $D_1^{**}$ (sub- or supercritical) which may belong either to one family  $\hat{L}_+$  
($\alpha \beta$ where $D_1^*$ is the same as $D_1^+$ and $D_1^{**}$ is the same is $D_1^+$) or to two different families $\hat{L}_+$  and $\hat{L}_-$ ($\alpha\beta^{'}$ where $D_1^*$ is the same as $D_1^+$ and $D_1^{**}$ is the same as $D_1^-$).  In the former case, the nontrivial  
motile branch has a turning point at a finite value of  $\P<1/4$ giving rise to a re-entrant behavior.  Similar behavior was also observed in some other nonlocal models \citep[e.g.][]{Kruse2003a, Tjhung2012, Giomi2014}.

\begin{figure}[!h]
\centering
\includegraphics[scale=0.6]{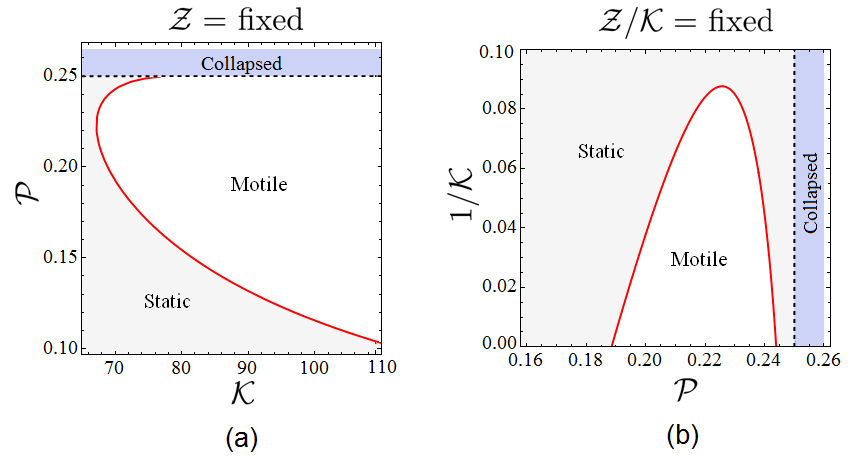}
\caption{\label{Adhesion} (a) Phase diagram of the system \eqref{eq.1} in the parameter plane $(\K,\P)$ at fixed $\Z=1$. (b) Phase diagram of the same system \eqref{eq.1} in the parameter plane $(\P,1/\K)$ at fixed $\Z/\K=0.015$. The solid (red) line indicates the motile bifurcation point ($D_1^+$ similar to Fig.~\ref{bifur1}), while the  black dashed lines  indicate the collapse threshold ($\P_{\text{max}}=1/4$) .}
\end{figure}

As illustrated in Fig.~\ref{bifur_P} and  shown more clearly in a phase diagram in Fig.~\ref{Adhesion}(a), in the re-entrant regime (sufficiently low $\K$), the increase of the average concentration of myosin (increase of $\P$ at fixed $\K$) first polarizes the cell and initiates motility, 
but then, if the contractility is increased further, the cell may becomes symmetric again by  re-stabilizing in
another static homogeneous configuration (see Fig.~\ref{bifur_P},  $\mathcal{K}=70$). We reiterate that re-symmetrization and arrest prior to  division (known also as `mitotic cell rounding') is a common feature of almost all animal cells \citep{stewart2011hydrostatic, Lancaster2013,Lancaster2014}.  In this respect,  it is interesting that if contractility ($\P$)  is increased further, the cell  collapses to a point because our effective  `size preserving spring'  cannot support the contraction any more.  Following \cite{Turlier2014}, we can associate such collapse with cell division. We can then argue that our deliberately minimalistic  model  succeeds in reproducing a rather general pattern of cell behavior by showing that symmetrization  (stabilization)  in space immediately precedes  the  division.

While the physical meaning of the non-dimensional parameter $\P$ in this discussion is rather clear (contractility measure), the significance of varying $\K$ at fixed $\Z$ is less obvious because both of these parameters depend on frictional  strength  of the background. Adhesivity of the cell to the substrate is  known to be a crucial parameter for motility initiation and arrest for various cell types \citep{Banerjee2011, Lober2014}.  To explicitly expose the role of friction, it is instructive  to interpret parameter $1/\K$ as a measure of  adhesivity  while maintaining  at a constant level the parameter $\Z/\K$, which   does not have any relation to surface friction and is just a ratio of diffusive and viscous time scales. 

The resulting phase diagram   in the $(\P,1/\K)$ plane at fixed $\Z/\K$ is shown in Fig.~\ref{Adhesion}(b).  In this diagram a horizontal path extending from left to right  means fixed adhesivity and increasing contractility.  One can see that at high adhesivity motility regimes cease to exist and static solutions  collapse as contractility increases. If the adhesivity is below a certain threshold,  the contractility increase  first causes polarization of a static configuration and motility initiation; further increase of contractility  causes re-symmetrization, arrest and eventually collapse.  An interesting regime corresponds to the very tip of the motile domain shown in Fig.~\ref{Adhesion}(b). Near this `critical' point the motility can be sustained  in a narrow `homeostatic window'  of parameters  and can be easily arrested by either increase or decrease of contractility. 

Very recently  new experimental results elucidating motility initiation in fish keratocytes have appeared \citep{Barnhart2015}.  According to these experiments, at a fixed contractility level (fixed $\P$ in our model),  the increase of  surface adhesivity (increase of $1/\K$ in our model) promotes static configurations while lowering adhesivity initiates motility. As it follows from Fig.~\ref{Adhesion}(b), these observations are in  agreement with our predictions. Our model also explains another observation made in \cite{Barnhart2015} that  at a fixed adhesivity, a blebbstatin (a contractility inhibitor) treatment promotes arrest of the  cells while a calyculin A treatment (a contractility stimulator) initiates motility.  The question whether a more substantial increase of contractility in experiment can lead to  re-symmetrization and arrest remains open. It is promising in this respect that some  cells are known to undergo  static to  motile transformation  in response to  a \emph{decrease} in the level of contractility \cite{Liu2010, Hur2011}. The  minimal model presented in \cite{Barnhart2015} is exactly  a 2D version of  the one formulated in \cite{Recho2013} and further developed in the present paper. While active protrusion  and  non linear  regulation of adhesion were also accounted for in \cite{Barnhart2015} to get a realistic cell shape, it is rather remarkable that the fundamental pattern of motility initiation  (including its dependence on contractility and adhesivity)  can be already captured within our  much more transparent setting,  see  Fig.~\ref{Adhesion}(b) and  Section \ref{experiment}.

\paragraph*{Nonlinear active stress} The fact that the bifurcation leading to polarization and motility initiation is always a supercritical pitchfork  indicates that  this model does not allow for metastability resulting in the  coexistence of motile and non-motile configurations that was observed in other models \citep[e.g.][]{Ziebert2013, Tjhung2012, Giomi2014}. To  obtain such  a coexistence in the present setting, we need to modify our model only slightly.  The main idea is to consider  a more realistic nonlinear dependence of the active stress on motor concentration   which is linear for small values of $c$ but then saturates after around a threshold  $c^*$. More specifically, we rewrite the main system of equations in the form 
\begin{equation}\label{BVPsaturation}
\begin{array}{rl}
-\Z\partial_{xx}\sigma+\sigma&=\P\Phi(r c)/r ,\\
\partial_t c+\K\partial_x(c\partial_x\sigma)&=\partial_{xx}c ,
\end{array}
\end{equation}
where, following \cite{Bois2011},  we choose a particular form of nonlinearity   $\Phi(x)= x/(1+x)$ and where we introduce the new non dimensional parameter $r=c_0/c^*$.

For simplicity we analyze below only the `rigid' limit when $k\rightarrow \infty$,  $L\rightarrow L_0$ while the  stress on the boundaries $-k(L/L_0-1)$ remains finite.  Notice that in this limit, which formally means that $\P \rightarrow  0 $ and $\K  \rightarrow  \infty $,  we have to re-scale the stress by $c_0\chi$ instead of $k$ and as a consequence in the rest of the section we denote,  with some abuse of notations,   $\sigma:=\sigma/\P$. The new dimensionless parameter replacing $\K$ and $\P$ is  
$
\lambda=c_0\chi/(\xi D)=\K\P
$ that is assumed to be finite \citep[see also][]{Bois2011, Howard2011, Hawkins2009a, Hawkins2011}.   In dimensionless variables the residual stress can be written as
$
\sigma_0=-\lim_{\mathcal{P}\rightarrow 0}\lim_{L\rightarrow 1}(L-1)/\P 
$.
\begin{figure}
\centering
\subfigure[ $r=1$]{
\includegraphics[scale=0.25]{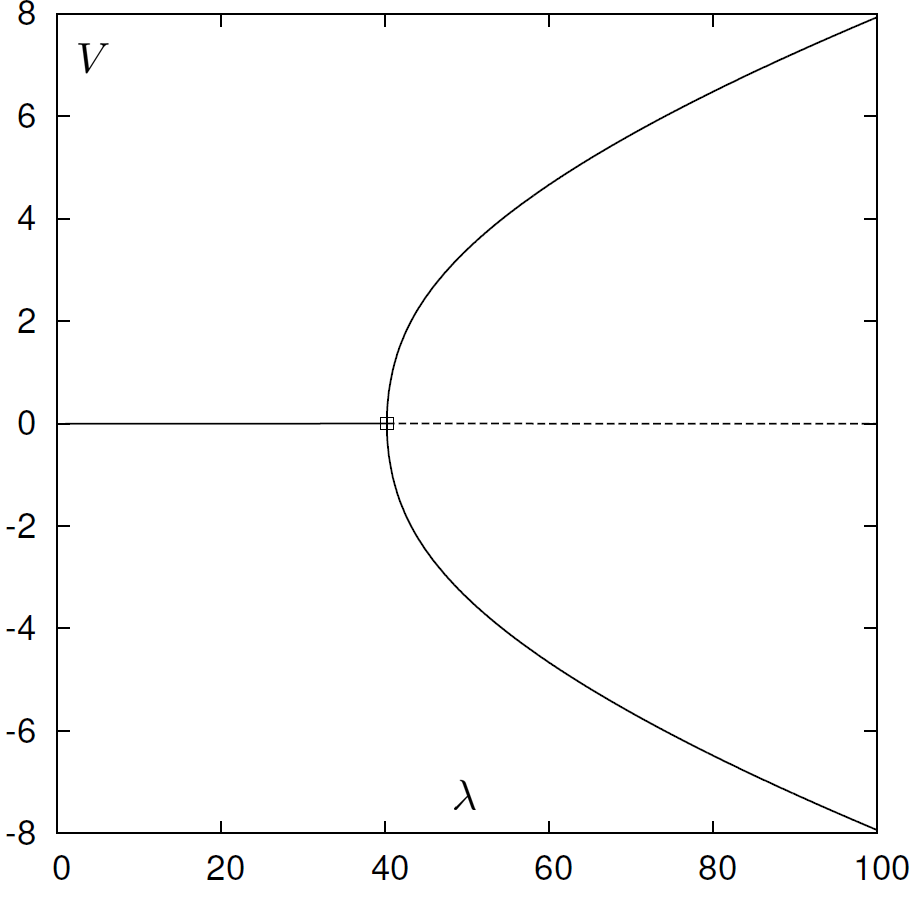}}
\subfigure[ $r=5$]{
\includegraphics[scale=0.25]{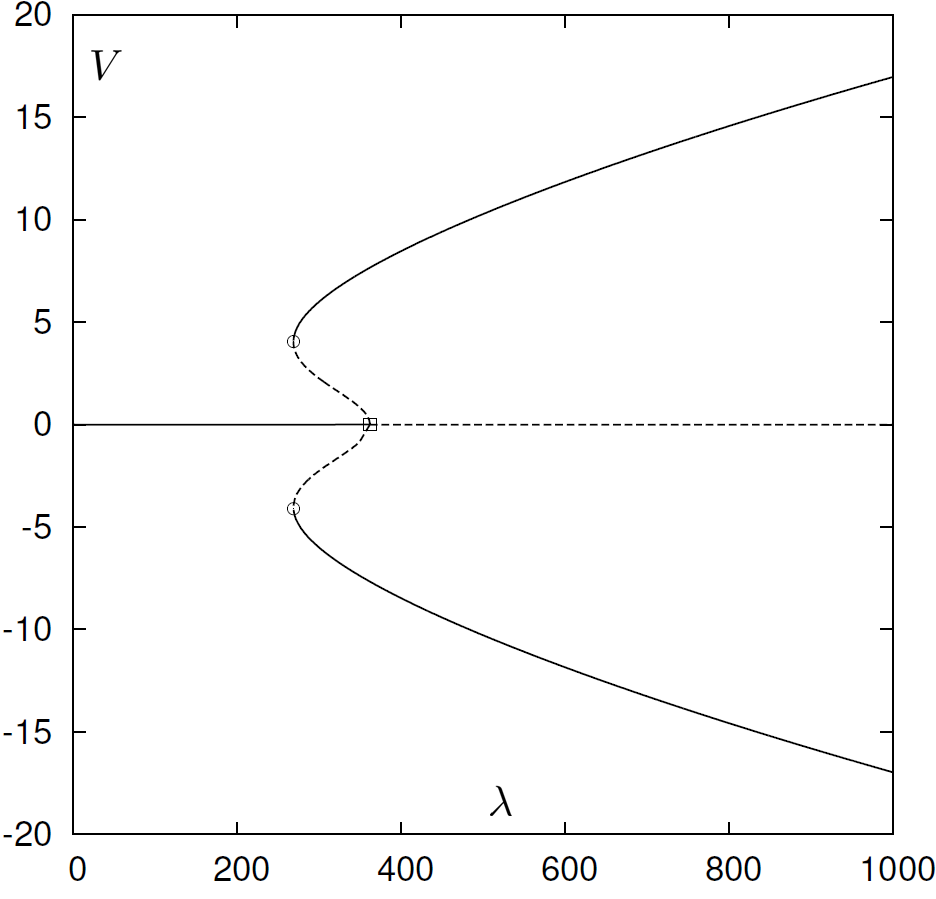}}
\caption{Bifurcation diagrams in the nononlinear model with fixed length (infinite stiffness)  (\ref{eqnstiff}) showing the possibility of a switch from supercritical to subcritical bifurcation.
Parameters: $\Z=1$.} \label{f:rtks_hds_LBD_VS}
\end{figure}
Then the boundary conditions read
$$
\begin{array}{rl}
\dot{l}_+-\dot{l}_-&=0\\
\sigma(l_{\pm}(t),t)&=\sigma_0\\
\partial_{x}c(l_{\pm}(t),t)&=0\\
\dot{l}_{\pm}&=\lambda\partial_{x}\sigma(l_{\pm}(t),t) .
\end{array} 
$$
For  TW solutions we can write the analogue of \eqref{eq.2} 
\begin{equation}\label{eqnstiff}
- \Z s^{''}+ s+s_0= \frac{\lambda}{r}\Phi\left(r \frac{\exp(s-Vu)}{\int_0^1 \exp(s-Vu)du}\right) ,
\end{equation}
where $s=\lambda (\sigma-\sigma_0)$ and $s_0=\lambda \sigma_0 $. The boundary conditions take the form $ s(0)=s(1)=0$ and  $s'(0)=s'(1)=V$.
The difference with our static solutions, described in Section \ref{Sectionstatconf}, is that
now we have to find the stress at the boundary $s_0$  instead of the length $L$. 

The analysis of the motility initiation bifurcation in this case  is presented in Appendix D.  The results are illustrated in Fig.~\ref{f:rtks_hds_LBD_VS}. As we see, when the nondimensional parameter $r$ is small, which means that we are in the linear regime, the bifurcation from static to motile regime is a supercritical pitchfork. However, at larger values of $r$ the nature of the bifurcation changes from supercritical to subcritical. This creates  a domain of parameters where static and motile regime can coexist and where the system may exhibit metastability and hysteresis.  Another important effect is that in this range of parameters the motility initiation/arrest  is a discontinuous transition which may explain why experimenters were unable to observe particularly small velocities of self propulsion in keratocytes \citep{Barnhart2011}. An alternative explanation of this experimental fact based on the idea of optimality and compatible with the supercritical nature of the motility initiation bifurcation, was proposed by \cite{Recho2014}

\section{Stability of  post-bifurcational regimes }\label{Numericalstudy}

Stability of various branches of the TW solutions identified in the previous sections was studied numerically. Since we have to deal with a moving segment, it is convenient  to map  system \eqref{eq.1} onto the fixed domain $[0,1]$ which makes the coefficients of the governing equations time dependent. To this end, we introduce the new space variable
$u(x,t)=[x-l_-(t)]/L(t)\in[0,1]$ and denote the new unknown functions $\hat{\sigma}(u,t)=\sigma[l_-+L(t)u,t]$ and $\hat{c}(u,t)=L(t)c[l_-+L(t)u,t]$. 
Then the original problem \eqref{eq.1}, \eqref{BCstress}--\eqref{BCcon} takes the form
\begin{equation}\label{unsteadynorm}
 -\frac{\Z}{L^2}\partial_{uu}\hat{\sigma}+\hat{\sigma} = \frac{\P}{L}\hat{c} 
\quad\text{and}\quad
 \partial_t \hat{c}+\frac{1}{L}\partial_u(\hat{v}\hat{c}) = \frac{1}{L^2}\partial_{uu}\hat{c},
\end{equation}
Here we defined the relative velocity $\hat{v}:=\K\partial_u\hat{\sigma}/L-\dot{G}-(u-1/2)\dot{L}$, where 
\begin{equation}
\begin{array}{rl}
\dot{G}&=(\K/L)\,[\partial_u\hat{\sigma}(1,t)+\partial_u\hat{\sigma}(0,t)]/2 , \\
\dot{L}&=(\K/L)\,[\partial_u\hat{\sigma}(1,t)-\partial_u\hat{\sigma}(0,t)] .
\end{array}
\end{equation}
The remaining boundary conditions can be written as 
\begin{equation}\label{unsteadynormbc}
\hat{\sigma}(u,t)=-(L-1) 
\quad\text{and}\quad
\partial_u\hat{c}(u,t) =0 \quad\text{at $u=\{0,1\}$.}
\end{equation}
while the  initial data take the form $\hat{c}(u,0)=\hat{c}^0(u),$ $G(0)=G^0$ and $L(0)=L^0$.

\begin{figure}
\centering
\includegraphics[scale=0.3]{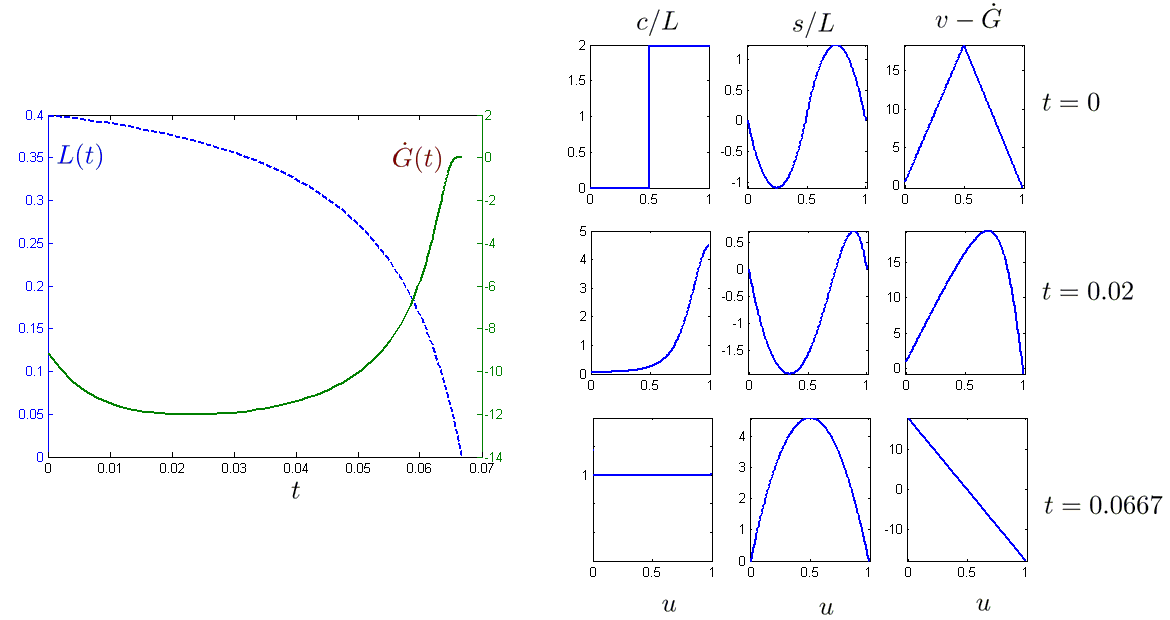}
\caption{\label{frommottosing} Cell length $L(t)$,  velocity $\dot{G}(t)$ and profiles $c/L(u,t)$, $s/L(u,t)$ and $v(u,t)-\dot{G}(t)$ 
for the test with initial data shown at $t=0$ with $L(0)=0.4$. Parameters $\P=0.245$, $\K=150$ and $\Z=1$ as in  Fig.~\ref{bifur_K}. The layer collapses  due to the contractile stress.}
\end{figure}
\begin{figure}
\centering
\includegraphics[scale=0.3]{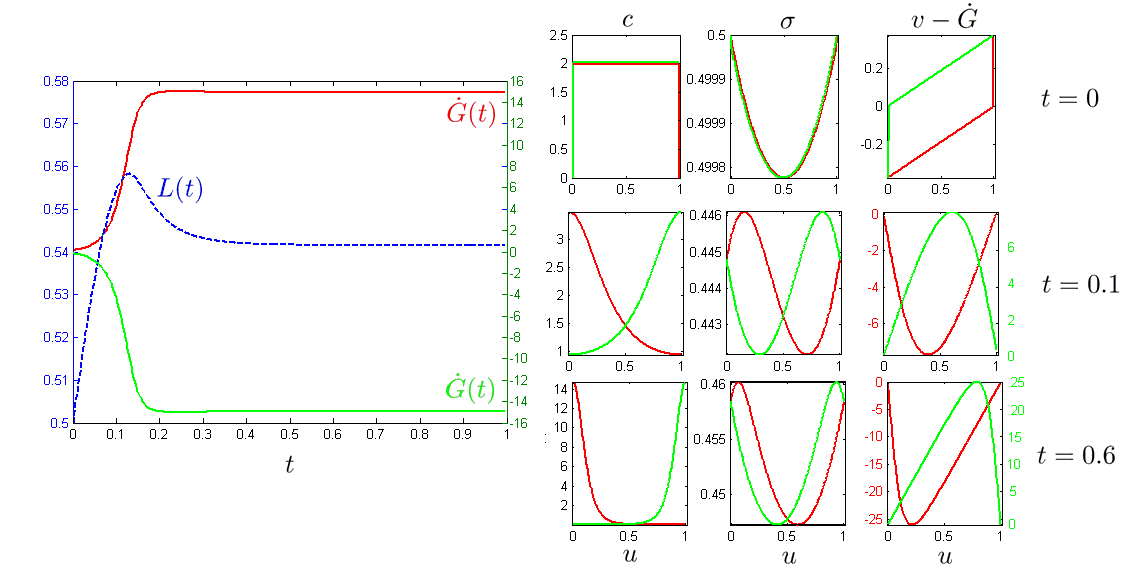}
\caption{\label{fromstattodyn} Cell length $L(t)$,  velocity $\dot{G}(t)$ and profiles $c$, $\sigma$ and $v(u,t)-\dot{G}(t)$ 
for the test with initial data shown at $t=0$ with $L(0)=0.5$. Parameters $\P=0.245$, $\K=150$ and $\Z=1$ as in  Fig.~\ref{bifur_K}. The layer polarizes to one the motile attractor (depending of the initial bias).}
\end{figure}
We  integrated  the dynamical system \eqref{unsteadynorm}--\eqref{unsteadynormbc} with initial data chosen close to one of the known steady states. The numerical scheme was based on the finite volume method \citep{Leveque2002}.  We used two dual regularly-spaced grids on the  interval $[0,1]$:  $Z$ and $Z_d$. Given the  initial condition  $\hat{c}$ we solved  \eqref{unsteadynorm}$_1$  on $Z$ and computed the effective drift term $\hat{v}$  on $Z_d$. We then applied the upwind finite volume scheme to \eqref{unsteadynorm}$_2$ and updated the concentration profile $\hat{c}$ on $Z$ which provided us with  the new initial
data for the next time step. The time interval for each time step was
adapted to ensure that the Courant-Friedrichs-Lewy condition is uniformly satisfied on $Z_d$.

Our numerical experiments suggest that the trivial branch $\hat{L}_-$ is unstable  together with all nontrivial non-singular static solutions. The singular static solutions from the $\hat{L}_0$ family appear to be locally stable. To illustrate the attractive nature of the singular static solutions
we choose in Fig.~\ref{frommottosing} the initial configuration with a length smaller than $\hat{L}_{-}$ with an internal initial profile biased to the front associated to a motile solution. We observe that the length collapses to zero in finite time and cell velocity goes to zero.  In accordance with the computations  made in Section \ref{Sectionstatconf}, the stress profile  converges to $s(u)/L\sim\K\P u(u-1)/2$, velocity to $v(u)\sim\K\P(u-1/2)$ and concentration to $c(u)=1$. 

\begin{figure}
\begin{center}
\includegraphics[scale=0.3]{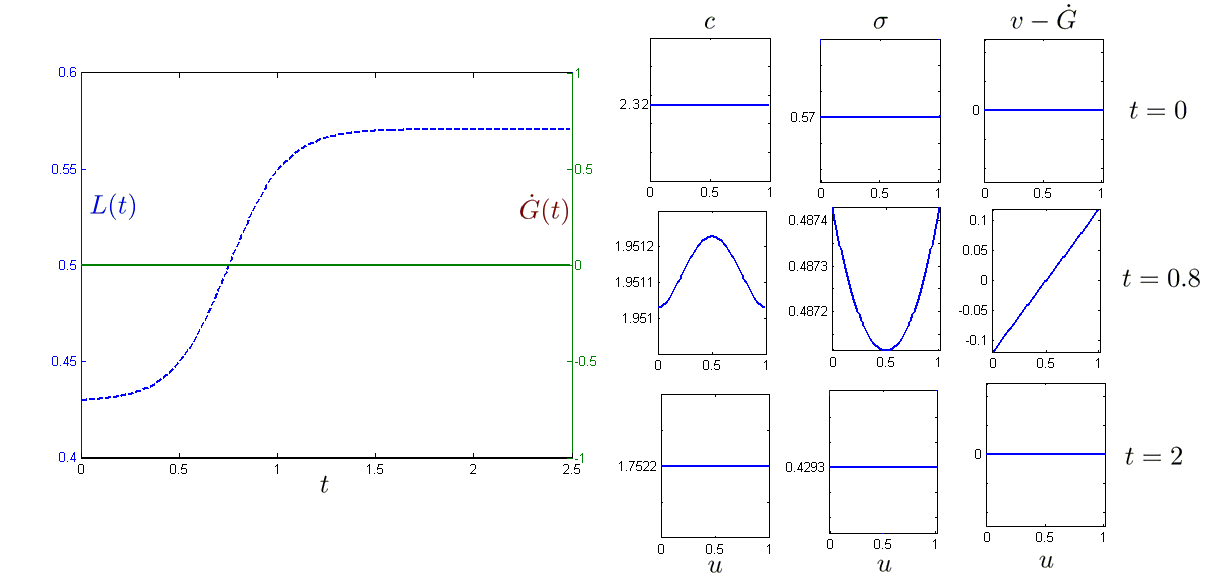}
\caption{\label{fromtrivtotriv} Cell length $L(t)$,  velocity $\dot{G}(t)$ and profiles $c$, $\sigma$ and $v(u,t)-\dot{G}(t)$ 
for the test with initial data shown at $t=0$ with  $L(0)=\hat{L}_-$ and the homogeneous concentration $c^0(u)=1/\hat{L}_-$.
Parameters $\P=0.245$, $\K=50$ and $\Z=1$ as in  Fig.~\ref{bifur_K}. The layer restabilises from the homogeneous branch $\hat{L}_-$ to $\hat{L}_+$.}
\end{center}
\end{figure}

Next, we observed numerically that the dynamic solutions are  all unstable except for  the branches bifurcating from the points $D_1^+$ on the trivial branch $\hat{L}_+$. The trivial branch $\hat{L}_+$ branch is locally stable until the first (motile) bifurcation $D_1^+$. Both symmetric subbranches of $D_1^+$
(subfamilies $1$ and $1^{'}$ in Fig.~\ref{bifur_K} and Fig.~\ref{profiles}) are stable. To illustrate the instability of a nontrivial static solution, we show in  Fig.~\ref{fromtrivtotriv} the escape of the phase trajectory from the neighborhood 
of the trivial static solution $\hat{L}_-$.  Since in this numerical test the value of $\K$ was chosen to be smaller than the critical value, corresponding to the bifurcation of the first motile branch $D_1^+$, 
the system originally placed near $\hat{L}_-$ becomes unstable and then re-equilibrates on another trivial static branch $\hat{L}_+$ without moving its geometrical center.

\begin{figure}[!h]
\begin{center}
\includegraphics[scale=0.5]{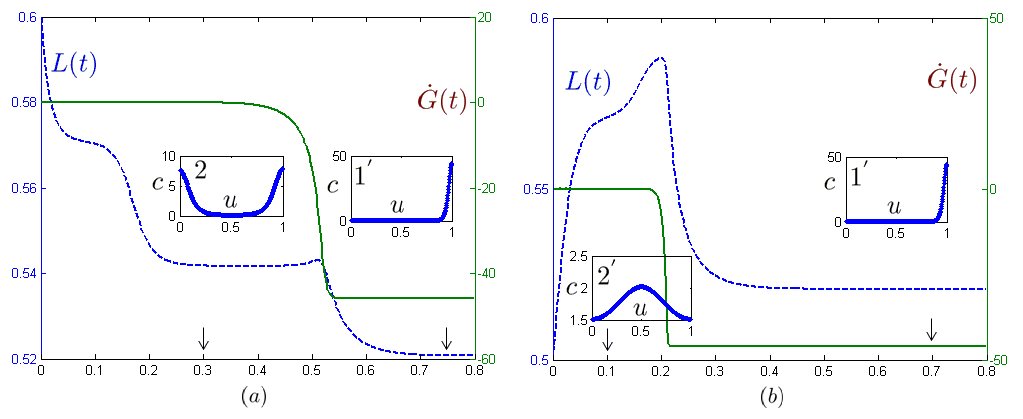}
\caption{\label{longliv}  Cell length $L(t)$ and velocity $\dot{G}(t)$ for the test with  $\P=0.245$, $\K=400$ 
and $\Z=1$ starting from homogeneous initial state with different initial lengths $L(0)=0.6$ (left) and $L(0)=0.5$ (right). 
The labels refer to Figs.~\ref{bifur_K}, \ref{profiles}. The two non trivial static branches bifurcating from $S_1^+$ denoted $2$ and $2'$ on Figs.~\ref{bifur_K} and Figs.~\ref{profiles}) have very different kinetic properties. }
\end{center}
\end{figure}

In Fig.~\ref{fromstattodyn} we illustrate motility initiation in two initially almost identical and nearly homogeneous static configurations which differ 
by a localized concentration peak  introduced either at the rear or at the front of the cell. We see that with time these two initial profiles converge to the 
different stable motile solutions $D_1$ and $D_1'$. The initial inhomogeneity is remembered and selects the subfamily of the $D_1$ solutions  with the same bias. As we see, independently of the direction of motion the cell recovers its length after a short  transient period. 

As in \cite{Bois2011, Howard2011, Kruse2003a}, who considered the problem with fixed boundaries, we find that some unstable multi-peaked static and dynamic solutions are long living. This behavior is reminiscent of the spinodal decomposition in a 1D Cahn-Hilliard model 
where the coarsening process gets critically slowed down near multiple saddle points \citep{Carr89}. To illustrate the long transients near the unstable   solutions we study in Fig.~\ref{longliv} evolution of two initially homogeneous concentration profiles with different initial length. We observe  that the phase trajectory first approaches the unstable branch from subfamily $2$ from Fig.~\ref{bifur_K} and Fig.~\ref{profiles}  before being finally attracted by the stable configuration from the subfamily $1'$.  Interestingly, the symmetric subfamily $2'$ can be also initially approached if we choose slightly different initial data, however, this regime is abandoned much faster than the solution from the subfamily $2$, see Fig.~\ref{longliv}(b). Based on our simulation, we conjecture that the lifespan of an unstable branch is linked to the distribution of motors and the states with higher localization of motors on the periphery of the cell survive longer than the states where motors are spread near the center of the cell. 
 
To summarize, we found considerable numerical evidence that in a problem with \emph{free} boundaries only trivial static solutions can be stable and only solutions with monotone profiles can describe  configurations of steadily moving cells. To confirm these results a more systematic mathematical analysis of stability of the obtained TW solutions is needed. Cells with \emph{constrained} or loaded boundaries may show different stability patterns as it is evidenced by the study of a related problem with a periodic boundary conditions \citep{Bois2011, Howard2011, Kruse2003a}.

\section{Mass transport of actin}\label{actinflow}

As we have already mentioned,  the   infinite compressibility assumption  allowed us to decouple the  force  balance equation from the mass balance equation.  Once the velocity field $v(x,t)=\K\partial_x\sigma(x,t)$ is known, the latter can be solved \emph{a posteriori } by the method of characteristics. 

Denote the trajectories of the mass particles  by $x=\phi(\zeta,t)$, 
where $l_-(0)\leq \zeta \leq l_+(0)$ is the Lagrangian coordinate   at $t=0$ and 
$l_-(t)\leq\phi(\zeta,t)\leq l_+(t)$.
The   characteristic curves can be found from the equations
\begin{equation}\label{e:phi_gal}
\frac{d \phi(\zeta,s)}{ds}=v(\phi(\zeta,s),s).
\end{equation}
Along these curves we must have
$$
\frac{d\rho(\phi( \zeta,s),s)}{ds}=-\rho( \phi( \zeta,s),s)\partial_xv(\phi(\zeta,s),s),
$$
Integration of this equation gives gives an explicit formula for the mass density 
\begin{equation}\label{densityeq}
\rho(\phi(\zeta,t),t)=\rho_0(\zeta)\exp\{-\int_0^t\partial_xv(\phi(\zeta,s),s)ds\}.
\end{equation}
As we are going to see below, this solution is applicable only outside  the singular points describing the sinks and the sources.  

Consider  a TW solution of \eqref{eq.1}
which satisfies the boundary conditions $l_-(t)=Vt$ and $l_+(t)=L+Vt$. Introducing the normalized co-moving variable 
$\hat{\phi}=(\phi-Vt)/L$ and the normalized Lagrangian variable $\hat{\zeta}=\zeta/L(0)$,
both in the interval $[0,1]$, we obtain that  $v=v(\hat{\phi})$ and Eq. (\ref{e:phi_gal}) reduces to 
\begin{equation}\label{44}
\frac{d\hat{\phi}(\hat{\zeta},t)}{dt}=\frac{v(\hat{\phi}(\hat{\zeta},t))-V}{L}.
\end{equation}
For TW solutions the general formula \eqref{densityeq}  describing the mass distribution  simplifies
\begin{equation}\label{rhoTW}
L\rho(\hat{\phi}(\hat{\zeta},t),t)\{v(\hat{\phi}(\hat{\zeta}))-V\}=L(0)\rho_0(\hat{\zeta})\{v(\hat{\zeta})-V\}.
\end{equation}

According to (\ref{44}) the points of the body where $v=V$ are singular 
because the relative flow there is stagnated. If at such point the slope of the function $v(\hat{\phi})$ is negative  
we obtain a sink of particle trajectories $\hat{\phi}=\gamma_+$ (i.e. an attractor for particles as $t\rightarrow \infty$)  
whereas if the slope of the function $v(\hat{\phi})$ is positive, the singular point  $\hat{\phi}=\gamma_-$ corresponds to a source of particle trajectories 
(an attractor as $t\rightarrow -\infty$).  An important feature of the flows described by (\ref{44}) is that  it takes an infinite time for a mass particle to reach a sink or to leave  a source because $(v(\hat{\phi})-V)^{-1}$ is not integrable in the neighborhood of $\gamma_{-}$ and $\gamma_{+}$.  
$$
\tau=\int_{\gamma_{-}}^{\gamma_{+}}\frac{d\hat{\phi}}{|v(\hat{\phi})-V|}=\infty.
$$
This implies that mass density infinitely localizes in the singular points (sources and sinks)  because
 $
L\rho(\hat{\phi})|v(\hat{\phi})-V|=\tau^{-1}=0 .
$
Then all mass points (corresponding to different  values of $\hat{\zeta}$)  come from the sources where the characteristic curves accumulate at large negative times and disappear in the sinks where the characteristic curves  accumulate at large positive time.  

For the trivial static solutions characterized by the lengths $\hat{L}_{\pm}$, 
there is no flow ($v-V=0$) and the mass density does not depend on either space or time. The density profiles for nontrivial static and motile solutions can be illustrated near the bifurcation points where the velocity profiles  are known explicitly. 

For instance, in the case of  the nontrivial static branches $S_m^{\pm}$ introduced in Section \ref{characteristicequation}, we obtain 
\begin{equation}\label{statbifur}
\frac{d\hat{\phi}(\hat{\zeta},t)}{dt}=\varsigma \sin(\omega_c\hat{\phi}(\hat{\zeta},t)),
\end{equation}
where $\omega_c=-2m\pi$. For determinacy, we choose the value of the amplitude $\varsigma$ in such a way that the maximum of our dimensionless velocity field is equal to one. The approximate value of $\varsigma$ can be computed  in the vicinity of the bifurcation point 
from  the amplitude equations  presented in Appendix C. In Fig.~\ref{anal1}  we show sample solutions of \eqref{statbifur} 
corresponding to homogeneous initial conditions $\hat{\phi}(\hat{\zeta},0)=\hat{\zeta}$ for positive and negative values of $\varsigma$ corresponding 
to the two possible branching directions. The corresponding density profiles are illustrated in Fig.~\ref{density} where the passive treadmilling cycles  are shown by arrows.  

Similarly, for the motile branches $D_m^{\pm}$ we need to solve the characteristic equation
\begin{equation}\label{dynabifur}
\frac{d\hat{\phi}(\hat{\zeta},t)}{dt}=\varsigma\left\{-\frac{L^2}{\omega_c^3\cos(\omega_c/2)}\left[ \omega_c \cos(\omega_c (\hat{\phi}(\hat{\zeta},t)-1/2))-2\sin(\omega_c/2)\right]-1 \right\},
\end{equation}
where $\omega_c$ is a solution of the equation \eqref{implicitomc2}.
Both equations can be solved analytically by separation of variables. In Fig.~\ref{anal2}, we show the sample solutions of  \eqref{dynabifur} 
corresponding to homogeneous initial conditions $\hat{\phi}(\hat{\zeta},0)=\hat{\zeta}$ again for the  positive and negative values of $\varsigma$. 

\begin{figure}
\centering
\subfigure[]{\includegraphics[scale=0.35]{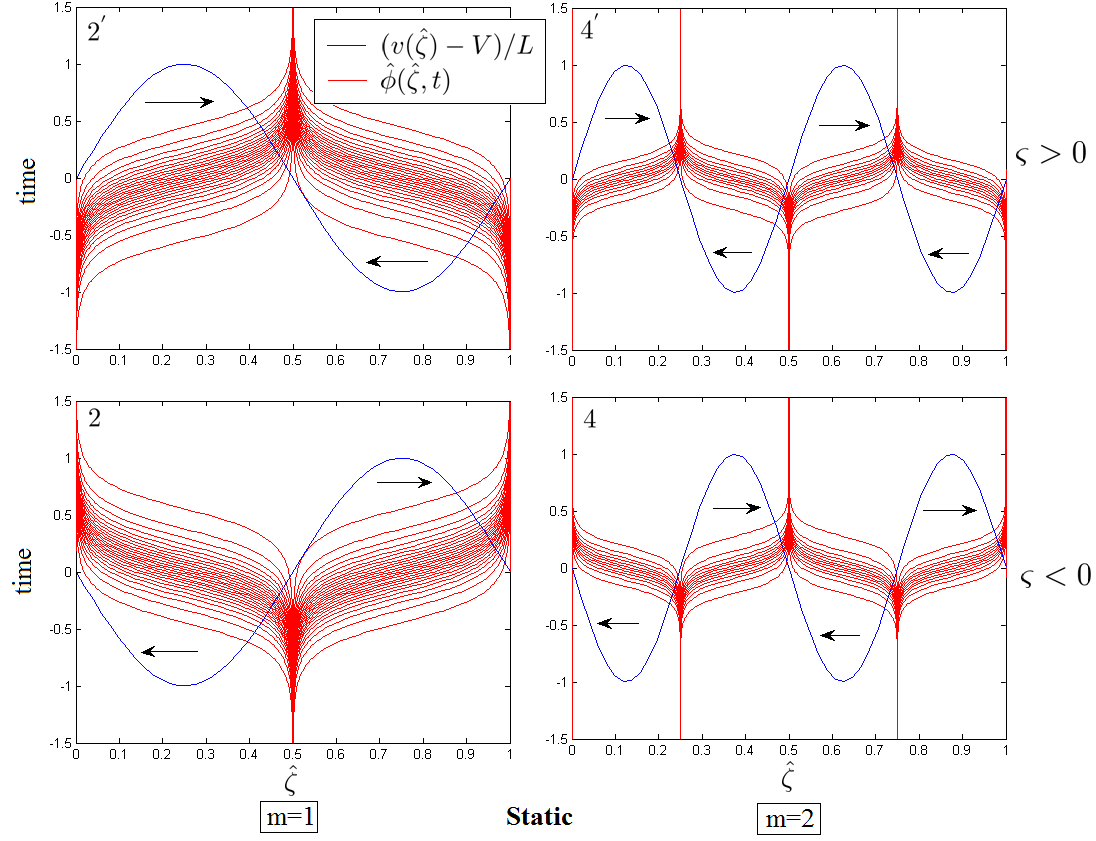}\label{anal1}}
\subfigure[]{\includegraphics[scale=0.35]{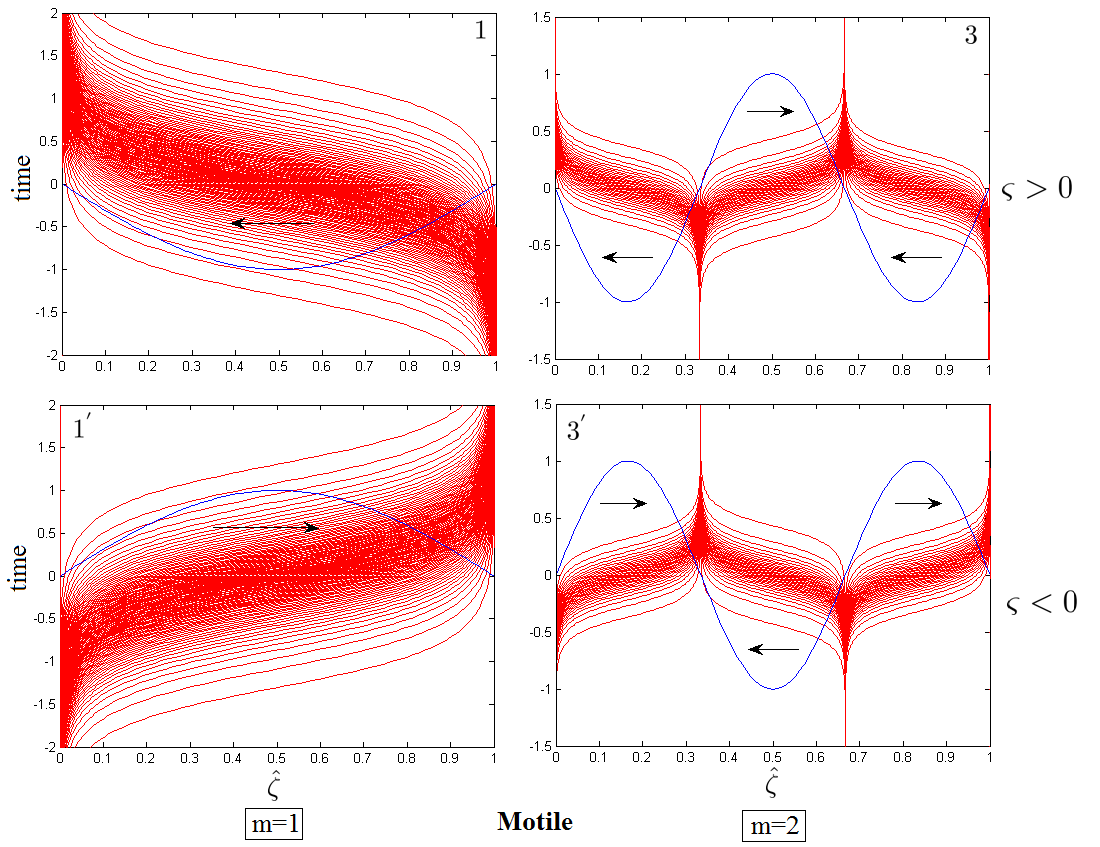}\label{anal2}}
\caption{(a) Trajectories of particles from sources to sinks for the first two static bifurcation points for initially homogeneously distributed set of particles. 
(b) Trajectories of particles from sources to sinks for the first two motile bifurcation points for initially homogeneously distributed set of particles. 
Labels $1, 1^{'}, 3, 3^{'}$ and labels $2, 2^{'}, 4, 4^{'}$ are related to Fig.~\ref{bifur_K} and Fig.~\ref{profiles}.
}
\end{figure}
\begin{figure}
\centering
\includegraphics[scale=0.4]{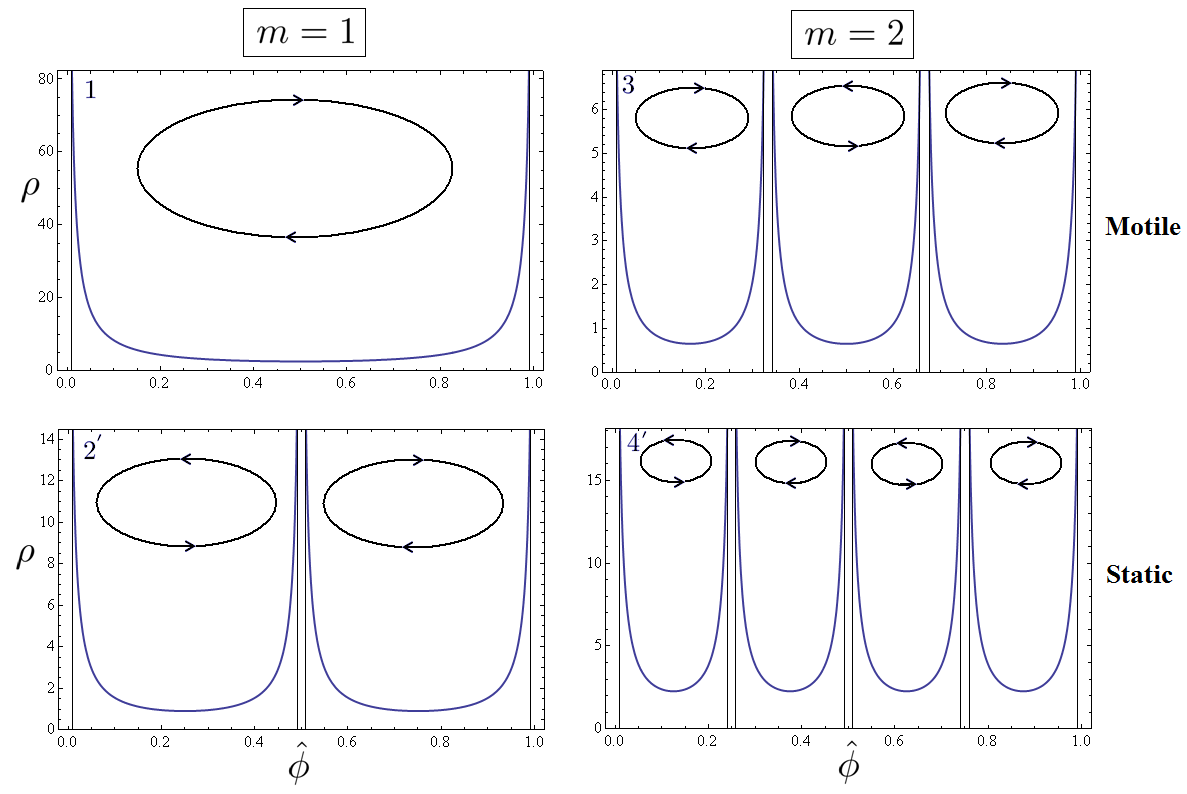}
\caption{\label{density} Density profiles for the first two motile and static branches for $\varsigma>0$, the profiles for $\varsigma<0$ are the same; only the treadmilling cycles (indicated by black circles) are going in the opposite direction. Labels are related to Fig.~\ref{bifur_K} and Fig.~\ref{profiles}. Parameter is $\epsilon=0.01$.}
\end{figure}
 
We reiterate that this model is singular because  in a one dimensional setting we are obliged to  over-schematize the treadmilling of actin.  
 
To recover the circulation aspect of the flow in a one-dimensional setting, we need  to regularize the problem near the singular  points and make the mass flux finite.  For instance,  we can cut out small regularizing 
domains of size $\epsilon$ around sinks and sources. 
In this way  we obtain  an effective `polymerization zone' around each source
$
\Gamma_{-}=\left\lbrace \hat{\phi}\in[0,1]/|\hat{\phi}-\gamma_{-}|<\epsilon \right\rbrace
$
and an effective `depolymerization zone' around each sink
$
\Gamma_{+}=\left\lbrace \hat{\phi}\in[0,1]/|\hat{\phi}-\gamma_{+}|<\epsilon \right\rbrace.
$ 
We assume that in the domain $\Gamma_{-}$ the  network is constantly assembled from the abundant monomers while in the domain $\Gamma_{+}$ it is constantly disassembled so that the pool of  monomers is  replenished. The ensuing  closure of the treadmilling cycle is instantaneous  (jump process) allowing the  monomers to avoid the  frictional contact with the environment. More precisely,  we assume that the jump part of the treadmilling cycle is a passive  equilibrium process driven exclusively by  myosin contraction.
The turnover time 
$$
\tau=\int_{\partial\Gamma_{-}}^{\partial\Gamma_{+}}\frac{d\hat{\phi}}{|v(\hat{\phi})-V|},
$$
is now finite and the corresponding density profiles are  illustrated  in Fig.~\ref{density}.  Notice that the flow 
between the neighboring  source and sink can be interpreted as a treadmilling cluster. Thus, for the $m^{\text{th}}$ static branch, 
we  have $2m$ such clusters and for the $m^{\text{th}}$ motile branch we have $2m-1$ clusters.   We reiterate that according to the numerical stability analysis conducted in Section \ref{Numericalstudy}, the only \emph{stable} motile solutions  as the ones with  a single treadmilling cycle extending from the leading to the trailing edge.

\section{Experimental verification of the model}\label{experiment}

\begin{figure}
\centering
\includegraphics[scale=0.75]{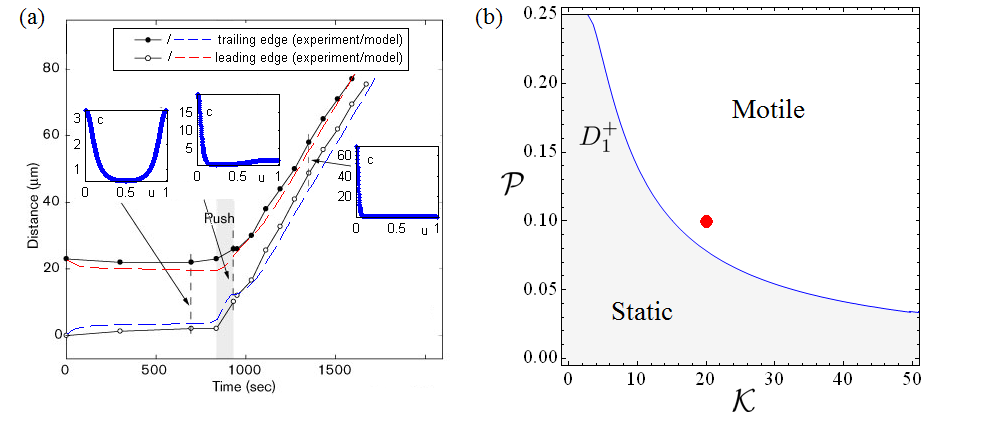}
\caption{\label{Verkhovsky} (a) Comparison of numerics with the experiments performed 
by Verkhovsky and co-authors in \cite{Verkhovsky1999}. 
Parameter values: $\Z=0.0125$, $\P=0.1$, $\K=20$. Integration is started from an initial cell length of $L_i=1.12$ with an homogeneous distribution of motors. In insets we show some snapshots of the distribution of motors obtained by numerical integration. The homogeneous configuration switches fast to a symmetric distribution where motors are relocating at both sides forming a two opposed lamellipods system. This state is long living but unstable and application of a transient loading leads to a break of symmetry and the subsequent localization of motors to the trailing edge forming a one lamellipod system. (b) Locus of the first motile bifurcation point associated to the homogeneous $\hat{L}_+$ branch for $\mathcal{Z}=0.0125$. Red dot shows the experimental data for keratocyte $\mathcal{P}=0.1$ and $\mathcal{K}=20$ which belongs to the motile regime. Such regime would be spontaneously reached under infinitesimal perturbations from a symmetric state but the long living nature of regime~$2$ (see Fig.~\ref{longliv}) makes it necessary to impose a transient asymetric perturbation to observe motility in experiments.}
\end{figure}

We can now compare the predictions of the model with experiments describing motility initiation in keratocytes. For instance, in the experiment of \cite{Verkhovsky1999},  a mechanical force was transiently applied via a micropipette on one side of a keratocyte fragment. Since the data presented in Fig.~5 of \cite{Verkhovsky1999} (and reproduced with permission in our Fig.~\ref{Verkhovsky})  are of  one dimensional nature we can directly  apply our model after adjusting it to account for mechanical loading. 

In order to make quantitative predictions we need to specify the values of parameters relevant for fish keratocytes. In \cite{Barnhart2011}, we find the values of viscosity $\eta \sim  10^{5} \, \text{Pa s}$ and active stress $\chi c_0 \sim 10^3 \,\text{Pa}$.  The drag coefficient can  vary over several orders of magnitude depending on the substrate whose physical properties have not been specified in \cite{Verkhovsky1999}.  However, based on the fact thati, in \cite{Verkhovsky1999}, the velocity of the fragment after initiation of motility was approximately $0.08\mu\text{m s}^{-1}$, we can infer from Fig.~5 of \cite{Barnhart2011} that $\xi \sim 2\times 10^{16}\,\text{Pa s m}^{-2}$. From \cite{Barnhart2011} and \cite{Luo2012}, we can also obtain the value of the diffusion coefficient $D \sim 0.25 \times 10^{-13}\,\text{m$^2$ s}^{-1}$ and, from \cite{Barnhart2010, Du2012, Loosley2012}, we estimate the stiffness of the cortex $k \sim 10^4\,\text{Pa}$. 
Finally, directly from \cite{Verkhovsky1999}, we infer that the characteristic length of the keratocyte fragment is $L_0 \sim 20\times 10^{-6}\,\text{m}$. 
Based on these estimates we conclude that $\Z \sim 0.0125$, $\P \sim 0.1$ and $\K \sim 20$.
\begin{figure}
\centering
\includegraphics[scale=0.5]{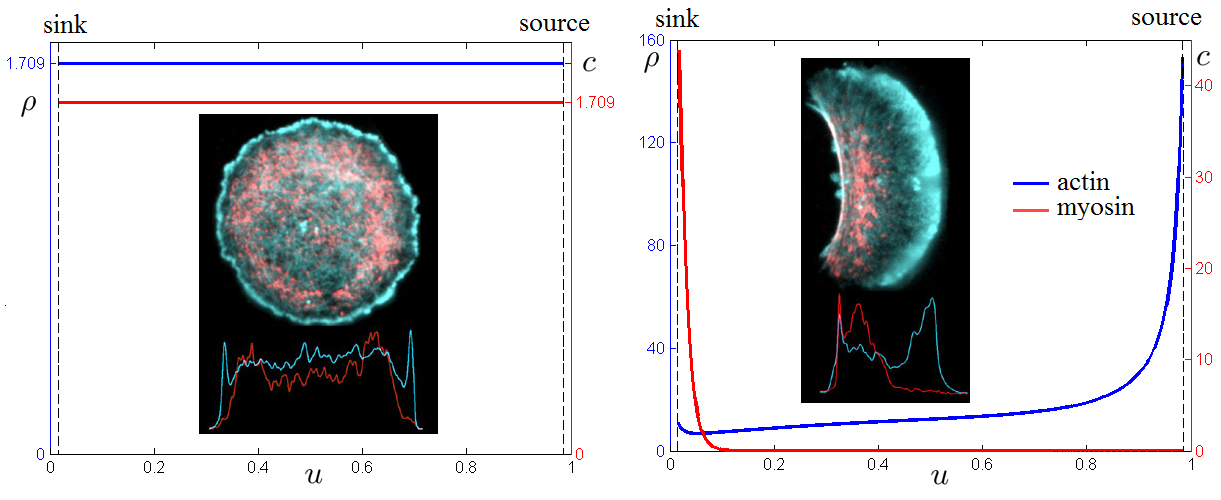}
\caption{\label{compaexpprofiles} Distribution of myosin (red) and actin (blue) in the static (left) and motile (right) regimes. Insets show the experimental distributions of actin (cyan) and myosin (red) from \cite{Verkhovsky1999}. Picture is taken from  http://lcb.epfl.ch/cms/lang/en/pid/71379, courtesy A. Verkhovsky. 
Parameter values: $\Z=0.0125$, $\P=0.1$, $\K=20$ and $\tau=0.018$.}
\end{figure}

In \cite{Verkhovsky1999} (Fig.~5), the  initially round fragment with  diameter $L_i=22$ $\mu$m, was subjected to an applied stress of the order of  $q_-=15-20$~kPa. The loading was applied after $830$~s and lasted  for about $80$~s.  The additional surface tractions can be easily incorporated into our model through the boundary condition at the rear of the cell:  
 $
\sigma(l_{-}(t),t)=-[L(t)/L_0-1]-q_-(t).
$

In Fig.~\ref{Verkhovsky}(a) we present the results of our numerical  simulation of the motility initiation experiment of Verkhovsky \citep{Verkhovsky1999}.  We start with a uniform initial  state where motors are distributed homogeneously. We chose a generic value of the  length $L(0)$ that is slightly different from the  value $\hat{L}_+$ which is unstable in this range of parameters.   The length first decays towards the  value corresponding to the branch $S_1^+$  as one could expect based on Figs.~\ref{bifur_K}, \ref{profiles} and \ref{longliv}. 
This is an unstable state which we found rather robust to selected   perturbations. The distribution of motors 
remains non polar with the development of two contractile zones characteristic of the nontrivial static regime $S_2^+$. The system then remains  in this long living unstable state until we apply 
an  additional one-sided force on the boundary breaking the  symmetry of the  $S_2^+$ state. The destabilized system evolves towards the motile state on the  $D_1^+$ branch with  both velocity and length well captured by our model. 

We can now compare with experiment  the  stationary density profiles (for both myosin and actin)  generated by the model. In the static regime, the flow of actin is absent ($v=0$) an  the model then predicts uniform distribution of actin and myosin. From Fig.~\ref{compaexpprofiles} (left), we see that  this prediction  is in agreement  with  experimental observations given that we disregard fluctuations and neglect near-membrane effects. 

From \cite{Rubinstein2009}, the turnover time of actin can be estimated to be $30$~s. Therefore we obtain in non dimensional units that $\tau=0.018$ which leads to the estimate $\epsilon=0.015$. We recall $\epsilon$ accounts for the size of polymerization source and sink at the leading and trailing edge, see Section \ref{actinflow} for details. Knowing  the value of~$\epsilon$, we can reconstruct the mass density distribution  $\rho(u)$ which we show  in Fig.~\ref{compaexpprofiles} (right) together with the motor concentration distribution  $c(u)$.  One can see that outside the boundary layers  the model captures the main effect: the sweeping of  actin towards the de-polymerization zone at the back of the cell by the retrograde flow and its regeneration on the polymerization zone at the front of the cell. A more detailed quantitative comparison with experiment requires an account of the two (or even three) dimensional nature of the flow.

Overall, we can conclude that the model reproduces rather well the  motility initiation pattern observed in Verkhovsky's experiment.  Moreover, the  
ensuing dynamics is described adequately by the stable motile branch predicted by our theory  formerly Fig.~\ref{Verkhovsky}(b).

In another  experiment by \cite{Yam2007}, which we interpret here only qualitatively because of the absence of a natural 1D representation, motility was induced by  injection of calyculin A,   known to be a factor increasing the activity of myosin motors. 
The conventional interpretation of this experiment refers to the local variation of contractility which disrupts the actin flow and affects the cascade of polymerization and depolymerization \citep{Paluch2006}. Instead, from the perspective of our model it is natural to conjecture that the injection calyculin A  affects  the value of parameter $\P$ pushing it beyond the threshold where the static symmetric configuration is stable and  initiating in this way the polarization instability which in turn leads to motility. 

Notice that in both experiments by \cite{Verkhovsky1999} and by \cite{Yam2007}, a fraction of keratocyte cells did not move at all 
 after being exposed to the same mechanical or chemical perturbation as the cells that did become motile. This can be explained by the fact that  the realistic  values for $\P$ and $\K$ lay rather close to the boundary 
separating static and motile regimes,  see Fig.~\ref{Verkhovsky}(b). It is then feasible that some cells remain in the symmetric (static) regime despite  the perturbation. 

It is also feasible that  the realistic  dependence of active stress on myosin concentration saturates above a certain threshold which, as we have seen,  can change the nature of the  motility initiation bifurcation  ($D_1$ branch)   into a \emph{subcritical} pitchfork, see discussion in Section \ref{characteristicequation}. This opens a finite range of  metastability  where both the homogeneous static state and the inhomogeneous motile state are locally stable.  The implied non-uniqueness  may be an alternative explanation of the simultaneous presence of motile and non motile cells despite apparently equal levels of contractility.
\begin{figure}
\centering
\includegraphics[scale=0.4]{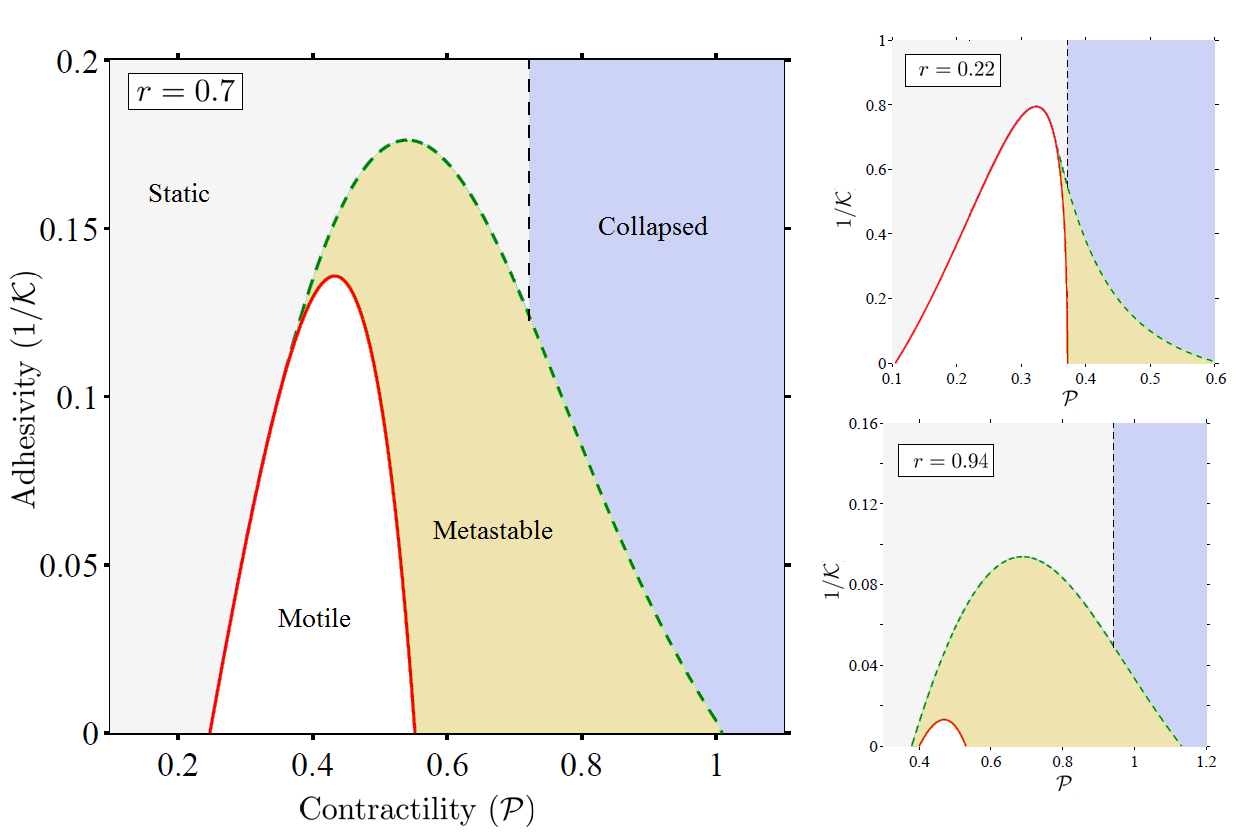}
\caption{\label{Phasediagadhesivity} Left: Phase diagram in the parameter plane $(\P,1/\K)$ for the system \eqref{BVPsaturation}  (with no length constraint).  The parameter $\Z/\K\approx 6\times 10^{-3}$ is fixed at its experimental value. The solid (red) line indicates the motile bifurcation threshold for the branch  $D_1^+$ (similar to Fig.~\ref{Adhesion}(b)), while the dashed line bounding the metastability domain  indicates the location of the turning points  on the motile branch in the appropriate analog  of Fig.~\ref{f:rtks_hds_LBD_VS}(b). The dashed line separating static and collapsed configurations indicates the location of the turning point $\alpha$  in Fig.~\ref{staticdiag}.  Right: effects of a high (top) and low (bottom) concentration saturation thresholds. }
\end{figure}

To exemplify this last claim, we show in Fig.~\ref{Phasediagadhesivity} the effect of switching to threshold type dependence of contractile stress on the concentration of motors, see \eqref{BVPsaturation}. Notice, that we have dropped in Fig.~\ref{Phasediagadhesivity} the assumption that the length of the moving segment is fixed. A comparison of Fig.~\ref{Phasediagadhesivity} with Fig.~\ref{Adhesion}(b) shows that the saturation of contractile stress introduces a finite zone of metastability between static and motile configurations: in this zone \emph{finite} perturbations are required to switch from static to a motile regime. This prediction was very recently confirmed in vivo by \cite{Barnhart2015} and the metastability domain as in Fig.~\ref{Phasediagadhesivity} was mapped experimentally.  
We also observe that for sufficiently large values of the active stress saturation threshold~$r$, our model associates metastability with both, motility initiation and motility arrest. On the arrest side \citep{Lancaster2014}, this prediction can be linked to the metastability of cell division \citep{Turlier2014} which to our knowledge has not been yet experimentally documented.

\section{Conclusions}

We studied  a prototypical model of a crawling segment of an  active gel   showing the possibility of spontaneous polarization and  steady self propulsion in the conditions when contraction is the only active process.  Our model, which focuses entirely on `pullers',  complements the existing theories of polarization and motility that place the main emphasis on `pushers' and link motility initiation  with active treadmilling and protrusion. Mathematically, the proposed model reduces to a dynamical system of Keller-Segel type, however, in contrast to its chemotaxic analog, the nonlocality in this model is due to \emph{mechanical} rather than \emph{chemical} feedback. If compared with previous studies of Keller-Segel type problems, our setting is complicated by the presence of free boundaries equipped with Stefan type boundary conditions.

As we argue, the motor proteins with sufficient contractility induce internal stress  which can overcome  the hydrodynamic resistance and induce flow. The  flow produces  a drift of motors in the direction of the regions where they  concentrate and such autocatalytic amplification is the mechanism of the  positive feedback in our model. The ensuing  runaway is countered by diffusion of motors which penalizes creation of concentration gradients and thus plays the role of a negative feedback. When a critical contractility of motors is reached, the homogeneous distribution of motors  becomes unstable. The
 contraction asymmetry  then induces a flow of actin filaments  towards the trailing edge thus producing frictional forces which propel the cell forward.  The rebuilding of the  balance between drift and diffusion  leads to the formation of  a  pattern. Among various admissible patterns, whose number increases with contractility, the stable ones localize motors at the trailing edge as observed in experiments.  

The proposed model provides an alternative \emph{qualitative} explanation of the experiments of \cite{Verkhovsky1999} and \cite{Yam2007} that have been previously interpreted in terms of active polymerization inducing the growth of  actin  network  \citep{Blanch-Mercader2013}. Most strikingly, the predictions of our model are also in \emph{quantitative} agreement with experimental data, which is rather remarkable in view of a schematic  nature of the model and the absence of fitting parameters. 

In addition, the model captures a durotactic effect since the directional motion cannot be initiated if friction with the substrate is larger than a threshold value. We show that below this threshold, motile regimes exist in a finite range of contractility. This means that if the cell is already in motion, it can recover the symmetric (static) configuration either by \emph{lowering} or by \emph{increasing} the amount of operating motors. The  predicted possibility  of cell arrest under the increased contractility should be  investigated in  focused  experiments. 

We have also shown that when the contractility depends on the motor concentration nonlinearly, one can have a metastability range where both static and motile regimes are stable and can coexist. In this range of parameters  a mechanical perturbation may be used to switch back and forth between static and dynamic regimes and  reproducing such behavior in vivo presents an interesting challenge.  This prediction of the model is particularly important in the context of collective cell motility (in tissues) where contact  interactions  are able to either initiate or terminate the motion \citep{Abercrombie1967,Heckman2009, Trepat2009, Vedula2012}.  

Despite the overall success of the proposed model, it leaves several important questions unanswered. Thus, our focus on a one dimensional  representation (projection)  of the motility process obscured the detailed description of the  reverse flow of actin monomers which we have replaced with an opaque  jump process.  Similarly, our desire to maximally  limit the number of  allowed activity  mechanisms,  forced us  to assume that  polymerization of actin monomers and their transport are  equilibrium processes. The assumption of infinite compressibility of the cytoskeleton, which is behind the decoupling of the mass transport  from the momentum balance,  also remains highly questionable   in the light of  recent advances in the understanding of cytoskeletal  constitutive response \citep{Broedersz2014, Pritchard2014}. Finally, our schematic depiction of focal adhesions  as passive frictional pads needs to be corrected by the account of the ATP driven integrin activity and the mechanical feedback from the binders to the cytoskeleton \citep{Schwarz2013}. 
These and other simplifying assumptions would have to be reconsidered  in a richer setting with realistic flow geometry which will also open a way towards more adequate description of  membrane (cortex) elasticity and will allow one to account for the polar nature of the gel \citep{Marchetti2013}.  

Ultimately, the  answer to the question whether   the proposed simplified mechanism provides  the \emph{fundamental} explanation of motility initiation in keratocytes will depend on the extent to which  the  inclusion of  all  other related factors affects the main conclusions of the paper.  A more thorough analysis  will also  open the way towards deeper understanding of  the remarkable \emph{efficiency} of the proposed mechanism  of self-propulsion delivering almost optimal performance at a minimal metabolic cost  \citep{Recho2014}.

\section{Acknowledgments}

P.R. acknowledges support from Monge Doctoral Fellowship at Ecole Polytechnique and a post doctoral grant from  Pierre Gilles De Gennes Foundation.
T.P. acknowledges support from the EPSRC Engineering Nonlinearity project EP/K003836/1. Authors thank D. Ambrosi, O. Du Roure, A. Grosberg, J.-F. Joanny and K. Kruse  for insightful comments.

\appendix 
\renewcommand\thefigure{\arabic{figure}}

\section{Analysis of nontrivial static solutions.} \label{A}

\begin{figure}[!h]
\centering
\includegraphics[scale=0.6]{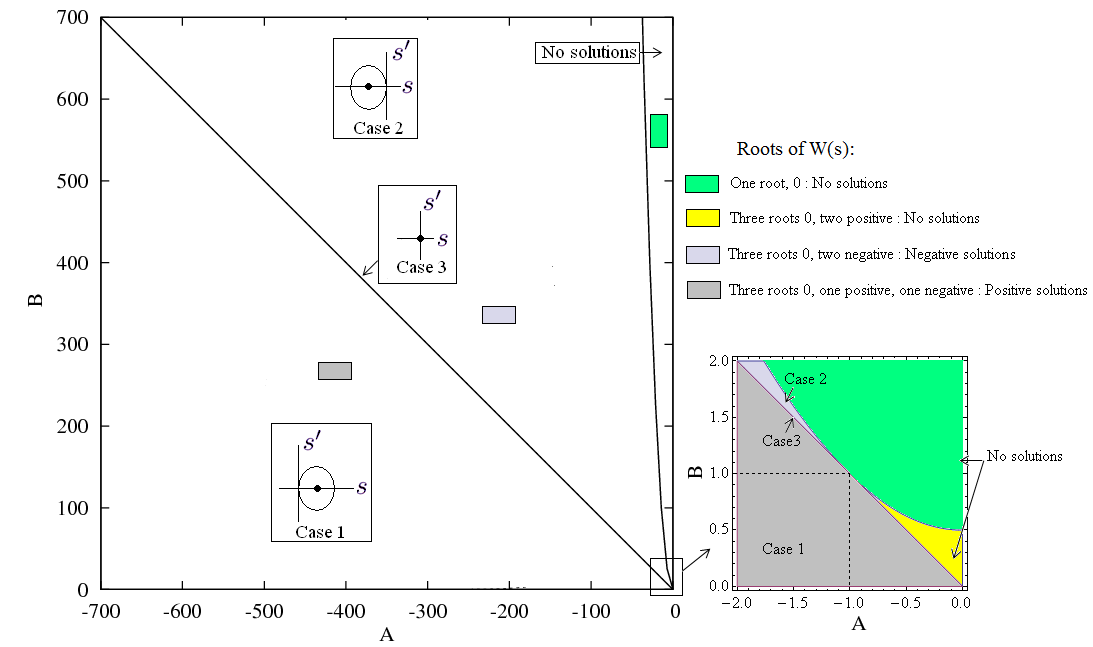}
\caption{\label{Conportrait} Phase diagram for the static solutions in the parameter space $(A,B)$. $A+B=0$ line is the trivial (homogeneous) solution. In the bottom  corner we show the blow up of the same diagram.}
\end{figure}

Solutions of boundary value problem \eqref{static} correspond to closed trajectories on the phase plane $(s,s')$ passing 
through the origin ($s=0,s'=0$) and different types of such trajectories are  illustrated in Fig.~\ref{Conportrait}.

Depending on the position of a point in the parametric plane $A,B$, one can identify five different types of behavior:
 
\begin{enumerate}
\item If $A+B=0$, then equation $W(r)=0$ has one double root at $r=0$ and one single root (negative or positive) at $r=s_-$ (Case 3 on Fig.~\ref{Conportrait}). 
The only solution is then the trivial one $s(u)=0$ and $L=\hat{L}_{\pm}$.
\item If $A+B<0$, then  Eq. $W(r)=0$  has three roots: $r=0$, $r=s_-<0$ and $r=s_+>0$.  This case corresponds to static branches labeled in Fig.~\ref{Conportrait}, Fig.~\ref{staticdiag}, Fig.~\ref{bifur_K} and Fig.~\ref{profiles} by numbers without a prime (Case 1 on Fig.~\ref{Conportrait}). In this domain we find nontrivial static solutions with $0\leq s(u)\leq s_+$. Different solutions correspond to different number ($m$) of sign changes for the function $s'(u)$ and different values of $L=2m\int_0^{s_{+}}W(\sigma)^{-\frac{1}{2}}d\sigma$.
\item If
 \begin{equation}\label{ineq2}
\left\{\begin{array}{c}
A+B>0\\
1-\sqrt{A^2-2 B+1}<B e^{-\sqrt{A^2-2 B+1}+A+1}\\
A<-1
\end{array}\right.
\end{equation}
then Eq. $W(r)=0$  has three roots: $r=0$ and $r=s_-<0$ and $r=s_+<0$ with, $s_+>s_-$.  This case corresponds to non motile branches labeled in Fig.~\ref{Conportrait}, Fig.~\ref{staticdiag}, Fig.~\ref{bifur_K} and Fig.~\ref{profiles}  by numbers with a prime~$'$ (Case 2 on Fig.~\ref{Conportrait}). In this domain, we find nontrivial static solutions with $s_-\leq s(u)\leq 0$. Again, different solutions correspond to different number of sign changes for the function $s'(u)$ and different values of $L=2m\int_0^{s_{+}}W(\sigma)^{-\frac{1}{2}}d\sigma$.
\item If
\begin{equation}\label{ineq3}
\left\{\begin{array}{c}
A+B>0\\
1-\sqrt{A^2-2 B+1}<B e^{-\sqrt{A^2-2 B+1}+A+1}\\
A>-1
\end{array}\right.
\end{equation}
then Eq. $W(r)=0$ has three roots: $r=0$ and $r=s_->0$ and $r=s_+>0$ with, $s_+>s_-$ and there are no static solutions since there are no closed paths in the phase plane passing through the point $(0,0)$.
\item If $1-\sqrt{A^2-2 B+1}>B e^{-\sqrt{A^2-2 B+1}+A+1}$, then equation $W(r)=0$ has only one non degenerate root at $u=0$. In this case there are no static solutions  since again there are no closed paths in the phase plane.
\end{enumerate}
Notice also that for the  solutions described above the map between the two parameterizations ($A,B$) and ($\K,\P$) is explicit 
\begin{equation}\label{map}
\begin{array}{c}
\mathcal{K}=A/(2m\int_0^{s_{+}}W(\sigma)^{-1/2}d\sigma-1)\\
\mathcal{KP}=2m\int_0^{s_{+}}(\sigma-A) W(\sigma)^{-1/2}d\sigma.
\end{array}
\end{equation} 

Notice also all nontrivial static solutions bifurcate from the trivial branches  in the sense that there are no detached solutions. Indeed, if a solution were detached, it would not pass through the origin (trivial solution) in the space $(s,s')$. But that would mean it cannot satisfies boundary conditions. 

\section{Analysis of the characteristic equation \eqref{33}.} \label{B}

Equation (\ref{33}) has two families of solutions depending on whether  $\omega$ is real or pure imaginary. In the first case, we denote $\omega_c\equiv|\omega|\geq 0$ whereas $\omega_c\equiv-|\omega|\leq 0$ in the second case. 
  
 \begin{equation}\label{bifurc_val}
 \begin{array}{rl}
 2[\cosh(\omega_c)-1]+(\Z\omega_c^2/\hat{L}^2-1)\omega_c\sinh(\omega_c) &=0 \quad\text{if $\omega_c^2>0$,} \\
 2[\cos(\omega_c)-1]+(\Z\omega_c^2/\hat{L}^2+1)\omega_c\sin(\omega_c) &=0 \quad\text{if $\omega_c^2<0$.}
 \end{array} 
 \end{equation}
Equations \eqref{bifurc_val}$_1$ and \eqref{bifurc_val}$_2$ need to be analyzed separately:
\begin{enumerate}
\item When $\omega_c^2>0$, equation \eqref{bifurc_val}$_1$ 
has a unique solution only if $\hat{L}^2/\Z\geq 12$. It is given by the implicit formula
$$2\tanh(\frac{\omega_c}{2})=(1-\frac{\mathcal{Z}}{\hat{L}^2}\omega_c^2)\omega_c.$$
The unstable  eigen-vector can be written as
$$
\begin{array}{rl}
\left(
\begin{array}{c}
  \delta L\\
  \delta V\\
  \delta s(u)
\end{array}
\right)
&= 
\left(
\begin{array}{c}
0\\
1\\
\frac{\hat{L}^2}{\Z\omega_c^3\cosh(\omega_c/2)}\big\{ \sinh[(u-1/2) \omega_c]-(2 u-1) \sinh(\omega_c/2)\big\}
\end{array}
\right)
\end{array}
$$
Since $\delta V\neq 0$, this bifurcation leads to a motile configuration  
which we denote $D_1$ . In Fig.~\ref{locus}   the eigen-functions associated with the sub-branch $D_1^{+}$ bifutrcating from the trivial solution $\hat{L}_{+}$  are illustrated for $\Z=0.01$. As parameter $\Z$ increases the exponential viscous boundary layers thicken. They fully disappear at $\Z=\hat{L}^2/ 12$ where the `hyperbolic' eigen-vectors become `trigonometric'.

\begin{figure}
\centering
\includegraphics[scale=0.3]{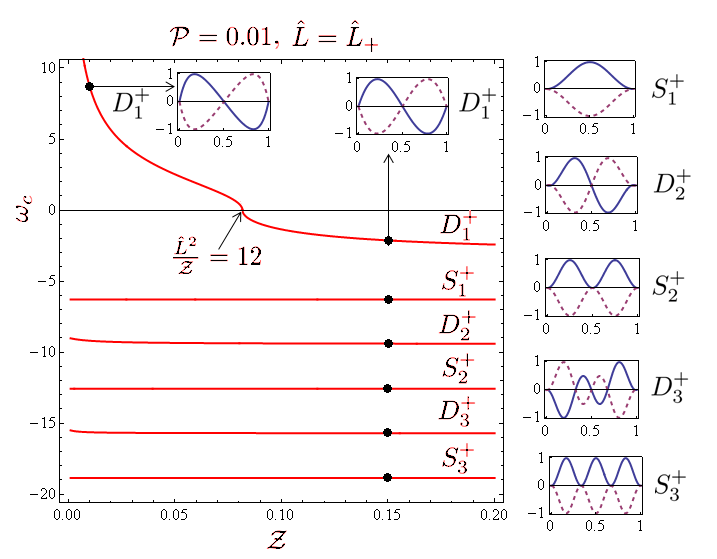}
\caption{\label{locus} Solution branches of the characteristic equation \eqref{bifurc_val} as functions of $\Z$ for
the trivial static solution $\hat{L}_+$ ($\P=0.01$).
From \eqref{e:omega_LKP}, the locus of the bifurcation points are recovered and shown of Fig.~\ref{bifur1}.
We refer to Fig.~\ref{bifur_K} for the label of bifurcation points. We represent in inserts the eigenfunctions  $\delta s$ related to $D_1^+,S_1^+,D_2^+,S_2^+,D_3^+,S_3^+$ for $\Z=0.15$ and the eigenfunction $\delta s$ related to $D_1^+$ for $\Z=0.01$. The eigenfunctions are normalized to $1$; solid and dashed lines correspond to the two possible directions of the pitchfork bifurcation.}
\end{figure}

\item When $\omega_c^2<0$, equation \eqref{bifurc_val}$_2$ has two sub-families of solutions:
\begin{enumerate}
\item The first family can be written explicitly :  $\omega_c=-2m\pi$ with $m\geq 1$. The unstable eigen-vector has the form
$$
\left(
\begin{array}{c}
  \delta L\\
  \delta V\\
  \delta s(u)
\end{array}
\right)
= \left(
\begin{array}{c}
1\\
0\\
\frac{(2\hat{L}-1)(\Z\omega_c^2+\hat{L}^2)}{\hat{L}^3(\hat{L}-1)}\big[\cos(\omega_c u)-1\big]
\end{array}
\right)
$$
Since $\delta V=0$, the bifurcated branch describes the nontrivial  static solutions $S_m$ studied in Section \ref{Sectionstatconf}. 
In Fig.~\ref{locus}  the eigen-functions associated with the sub-branch $S_m^{+}$ originating at the trivial solution $\hat{L}_{+}$
are illustrated for $\Z=0.15$. 
\item The second family consists of a countable set of negative roots of equation \eqref{bifurc_val}$_2$ given implicitly by,
\begin{equation}\label{implicitomc2}
2\tan(\frac{\omega_c}{2})=(\frac{\mathcal{Z}}{\hat{L}^2}\omega_c^2+1)\omega_c
\end{equation}
The unstable  eigen-vector is
$$
\begin{array}{rl}
\left(
\begin{array}{c}
  \delta L\\
  \delta V\\
  \delta s(u)
\end{array}
\right)
&=
\left(
\begin{array}{c}
0\\
1\\
\frac{-\hat{L}^2}{\mathcal{Z}\omega_c^3\cos(\omega_c/2)}\big\{ \sin[(u-1/2)\omega_c\big]-2(u-1/2)\sin(\omega_c/2)\big\}
\end{array}
\right)
\end{array}
$$
\end{enumerate}
It corresponds to motile branches because $\delta V\neq 0$. We denote this family by $D_m$. In Fig.~\ref{locus}  the eigen-functions associated with a subbranch $D_m^{+}$ originating at  trivial solutions $\hat{L}_{+}$ are illustrated for $\Z=0.01$.
\end{enumerate}

\section{Normal forms.} \label{C}

\paragraph*{Normal form in $\K$} In terms of the normalized stress  variable $r=s/L$ the original nonlinear problem can be written as
\begin{equation}\label{LSKeq}
-\Z r''(u)+L^2r(u)=\K\P\frac{e^{L(r(u)-Vu)}}{\int_0^1e^{L(r(u)-Vu)}du} ,
\quad\text{with}\quad
r(0)=r(1)=0, \quad r'(0)=r'(1)=V.
\end{equation}
Assume that  $\epsilon$ is a small parameter and expand the solution of \eqref{LSKeq} around a bifurcation point up to third order
$$
r=0+\epsilon\overset{1}{r}+\epsilon^2\overset{2}{r}/2+\epsilon^3\overset{3}{r}/6+o(\epsilon^3) , \quad 
V=0+\epsilon\overset{1}{V}+\epsilon^2\overset{2}{V}/2+\epsilon^3\overset{3}{V}/6+o(\epsilon^3) , \quad
L=\hat{L}+\epsilon\overset{1}{L}+\epsilon^2\overset{2}{L}/2+\epsilon^3\overset{3}{L}/6+o(\epsilon^3) .
$$
Assume that the  bifurcation parameter $\K$ and therefore 
$$
\K=\overset{0}{\K}+\epsilon\overset{1}{\K}+\epsilon^2\overset{2}{\K}/2+\epsilon^3\overset{3}{\K}/6+o(\epsilon^3).
$$
where $\overset{0}{\K}$ is the bifurcation point.  These expressions are then inserted into equation \eqref{LSKeq}. Separating different  orders of $\epsilon$ we obtain three differential equations
\begin{align}
O(1), \qquad \mathbb{L}(\overset{1}{r},\overset{1}{L},\overset{1}{V})&=0, \label{firstorder}\\
O(2), \qquad\mathbb{L}(\overset{2}{r},\overset{2}{L},\overset{2}{V})&=\overset{0}{\K}\,\mathbb{P}_0(\overset{1}{r},\overset{1}{L},\overset{1}{V})+\overset{1}{\K}\,\mathbb{P}_1(\overset{1}{r},\overset{1}{L},\overset{1}{V}),\label{secondorder}\\
O(3), \qquad\mathbb{L}(\overset{3}{r},\overset{3}{L},\overset{3}{V})&=\overset{0}{\K}\,\mathbb{Q}_0(\overset{1}{s},\overset{1}{L},\overset{1}{V},\overset{2}{r},\overset{2}{L},\overset{2}{V})+\overset{1}{\mathcal{K}}\,\mathbb{Q}_1(\overset{1}{r},\overset{1}{L},\overset{1}{V},\overset{2}{r},\overset{2}{L},\overset{2}{V})
+\overset{2}{\K}\,\mathbb{Q}_2(\overset{1}{r},\overset{1}{L},\overset{1}{V},\overset{2}{r},\overset{2}{L},\overset{2}{V}), \label{thirdorder}
\end{align}
where $\mathbb{L}$ is the linear operator already introduced in the stability analysis, see \eqref{bifurlineq}:
$$\mathbb{L}(r,L, V):=r''(u)-\omega^2  r(u)+\frac{\left(u-\frac{1}{2}\right) \left(\omega^2 Z-\hat{L}^2\right)}{Z}V+\frac{(2 \hat{L}-1) \omega^2 \left(\omega^2 Z-\hat{L}^2\right)}{(\hat{L}-1) \hat{L}^4}L$$
and $\mathbb{P}_0$, $\mathbb{P}_1$, $\mathbb{Q}_0$, $\mathbb{Q}_1$ and $\mathbb{Q}_2$ are known non linear operators.
The boundary conditions remain the same at all orders $i$:
$$\overset{i}{r}(0)=\overset{i}{r}(1)=0, \quad \overset{i}{r'}(0)=\overset{i}{r'}(1)=\overset{i}{V}$$

In the leading order, we obtain the results already reported in Section~\ref{characteristicequation} including  the eigenvalue $\overset{0}{\K}$ and the eigenfunction $\overset{1}{r}(u),\overset{1}{L},\overset{1}{V}$. To have a nontrivial solution in the next order, the right-hand side of equation \eqref{secondorder} must be  orthogonal to the kernel of the dual of $\mathbb{L}$ (for the $L^2$ scalar product). In  the $(C_1,C_2,\delta L,\delta V)$ space, see Section~\ref{characteristicequation}, 
this means orthogonality to the kernel of the transpose of matrix \eqref{matri}. The resulting linear scalar equation   determines the value of $\overset{1}{\K}$.  When this value vanishes, higher orders must be considered in a similar way. We summarize below the main results obtained by implementing this procedure. 
\begin{figure}
\begin{center}
\subfigure[]{\includegraphics[scale=0.3]{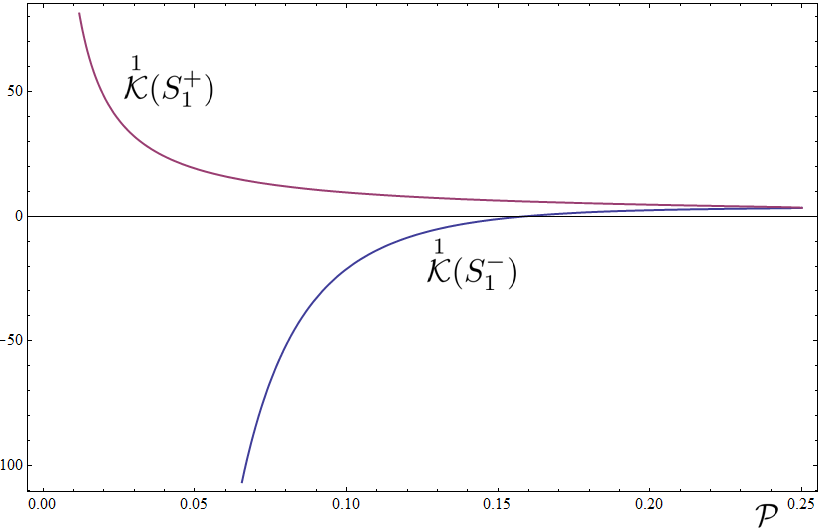}\label{subtosuppitch2}}
\subfigure[]{\includegraphics[scale=0.3]{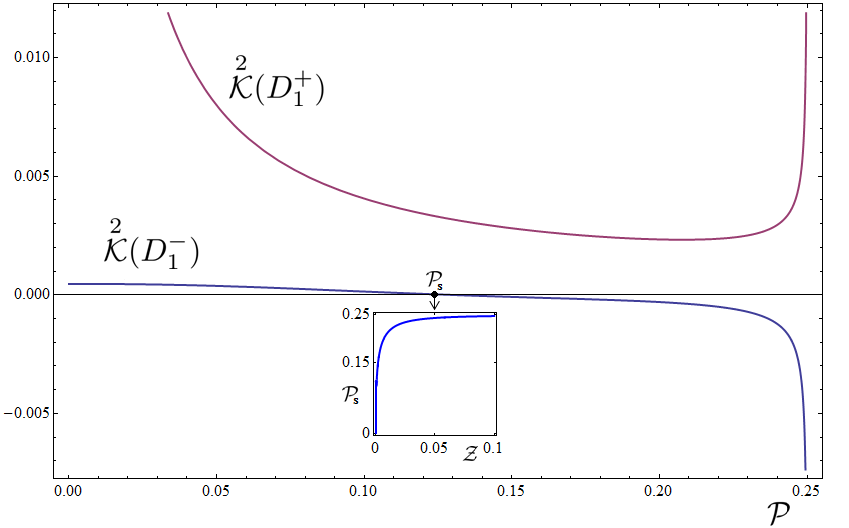}\label{subtosuppitch}}
\caption{
(a) Values of $\overset{1}{\mathcal{K}}$ for the $S_1^+$ and $S_1^-$ branches 
as functions of parameter $\P$ for $\Z=0.001$.
(b) Values of $\overset{2}{\mathcal{K}}$ for the $D_1^+$ and $D_1^-$ branches 
as functions of parameter $\P$ for $\Z=0.001$. The point where $\overset{2}{\K}=0$ along the $\hat{L}_-$ static 
branch indicates a nature change of the motile pitchfork bifurcation from supercritical to subcritical pitchfork.
The parameter dependence of such a point is represented as a function of $\P=\P_s(\Z)$ in the inset.} 
\end{center}
\end{figure}

\begin{enumerate}
\item \emph{Static branches} result from transcritical bifurcations. For the $m^{\text{th}}$ branch we have
$$ 
\overset{1}{\K}=(\hat{L}^2-4m^2\pi^2\Z)/[\hat{L}^3(\hat{L}-1)]\\
$$
\item \emph{Motile branches} all correspond to pitchfork bifurcations  with $\overset{1}{\K}=0$. 
They can be either subcritical or supercritical depending on the sign of 
{\footnotesize
\begin{multline}
\overset{2}{\K}=\left( (2 \hat{L}-1) \hat{L}^{14} \left(3 \omega^2+770\right)+2 \hat{L}^{12} \Z \left(3 (8-11 \hat{L}) \omega^4+1415 (1-2 \hat{L}) \omega^2+4620 (1-2 \hat{L})\right)+3 \hat{L}^{10} \omega^2 \Z^2 \left(40 \hat{L} \left(\omega^4+61
   \omega^2+\right. \right. \right. \\
   \left. \left. \left. 374\right)-31 \omega^4-1340 \omega^2-7480\right)+2 \hat{L}^8 \omega^4 \Z^3 \left(4 \hat{L} \left(6 \omega^4+89 \omega^2-3150\right)-9 \omega^4+50 \omega^2+7380\right)+\hat{L}^6 \omega^6 \Z^4 \left(-2 \hat{L} \left(165 \omega^4+\right. \right. \right.\\
    \left. \left. \left.6502 \omega^2+23160\right)+195
   \omega^4+6574 \omega^2+22440\right)+6 \hat{L}^4 \omega^8 \Z^5 \left((61 \hat{L}-34) \omega^4+(2622 \hat{L}-1339) \omega^2+7072 (2 \hat{L}-1)\right)+\right.\\
  \left. \hat{L}^2 \omega^{10} \Z^6 \left(3 (31-60 \hat{L}) \omega^4+4264 (1-2 \hat{L}) \omega^2-28224 (2\hat{L}-1)\right)+2 (2 \hat{L}-1) \omega^{12} \left(9 \omega^4+472 \omega^2+3456\right) \Z^7\right) \\
  \left( 144 (\hat{L}-1) (2 \hat{L}-1) \omega^8 \Z^3 \left(\hat{L}^2-\omega^2 \Z\right) \left(\hat{L}^4-2 \hat{L}^2 \left(\omega^2+6\right) \Z+\omega^4
   \Z^2\right)\right)^{-1}
\label{e:K2}
\end{multline}
}
\end{enumerate}

In Fig.~\ref{subtosuppitch} we illustrate the function $\overset{2}{\K}(\mathcal{P})$ 
for the first motile branches $D_1^-$ and $D_1^+$. As $\overset{2}{\mathcal{K}}\geq 0$ for all values of the activity parameter $\P$, the motile branch $D_1^+$ always bifurcates 
from the  static branch in a supercritical (pitchfork) manner. In contrast, the motile branch $D_1^-$  can 
bifurcate either supercritically or a subcritically depending on the value of  $\P$.
When $\P$ is larger than a threshold value $\P_s$, the coefficient $\overset{2}{\K}$ changes sign and becomes negative indicating 
a subcritical character of the bifurcation on the $\hat{L}_-$ static branch.

\paragraph*{Normal form in $\P$} We now consider $\mathcal{P}$ as the bifurcational parameter. The derivation of the normal form  in this case is more complex   because the homogeneous static solution $\hat{L}(\mathcal{P})$ is a multivalued function 
of $\P$ (see Fig.~\ref{bifur2}). 
One can circumvent the difficulty by introducing a new variable
\begin{equation}\label{mapPtoJ}
J=L(L-1)+\P,
\end{equation}
Then the boundary value problem (\ref{LSKeq}) takes the form
$$
-\Z r''(u)+L^2r(u)=\K[J-L(L-1)]\frac{e^{L(r(u)-Vu)}}{\int_0^1e^{L(r(u)-Vu)}du} ,
\quad\text{with}\quad
r(0)=r(1)=0, \quad r'(0)=r'(1)=V.
$$
whose trivial  solution is $(J,V,r)=(0,0,0)$. In this formulation 
  $J$, $V$ and $r(u)$ are unknowns while the length $L$ is the bifurcation parameter. 
The regular expansions near the homogeneous state give
$$
r=\epsilon\overset{1}{r}+\epsilon^2\overset{2}{r}/2+\epsilon^3\overset{3}{r}/6+o(\epsilon^3), \quad
V=\epsilon\overset{1}{V}+\epsilon^2\overset{2}{V}/2+\epsilon^3\overset{3}{V}/6+o(\epsilon^3), \quad
J=\epsilon\overset{1}{J}+\epsilon^2\overset{2}{J}/2+\epsilon^3\overset{3}{J}/6+o(\epsilon^3),
$$
$$
L=\overset{0}{L}+\epsilon\overset{1}{L}+\epsilon^2\overset{2}{L}/2+\epsilon^3\overset{3}{L}/6+o(\epsilon^3).
$$
Distinguishing the static and motile branches as before, we obtain the following results:
\begin{enumerate}
\item \emph{Static branches} are all found to be transcritical bifurcation. For the $m^{\text{th}}$ branch, we have  $\P=\overset{0}{\P}+\epsilon\overset{1}{\P}+o(\epsilon)$ where
$$
\overset{0}{\P}=-\overset{0}{L}(\overset{0}{L}-1) ,  \overset{1}{\P}=1-(2\overset{0}{L}-1)\overset{1}{L}.
$$
and
$\overset{0}{L}$   is a solution of the cubic equation
$$
-\K(\overset{0}{L})^2(\overset{0}{L}-1)=\Z4\pi ^2m^2+(\overset{0}{L})^2 .
$$
In this equation only two roots corresponding to $S_m^*$ (the smaller) and to $S_m^{**}$ (the larger) are in the range $[0,1]$.  
In Fig.~\ref{normformS1P}, we illustrate the behavior of of the function $\overset{1}{\P}(\K)$ for the branches $S_1^*$ and  $S_1^{**}$.
\begin{figure}
\begin{center}
\subfigure[]{\includegraphics[scale=0.4]{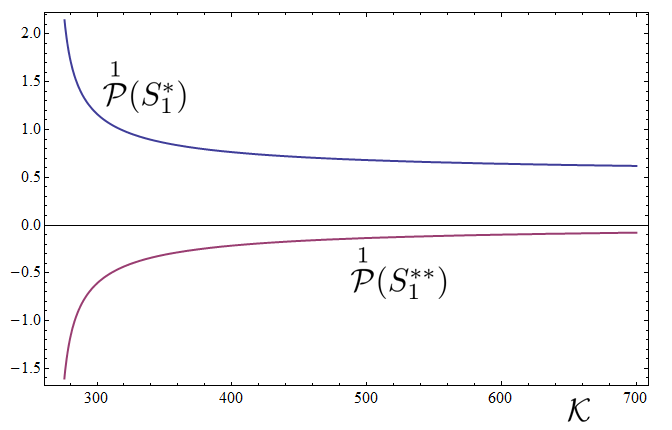}\label{normformS1P}}
\subfigure[]{\includegraphics[scale=0.29]{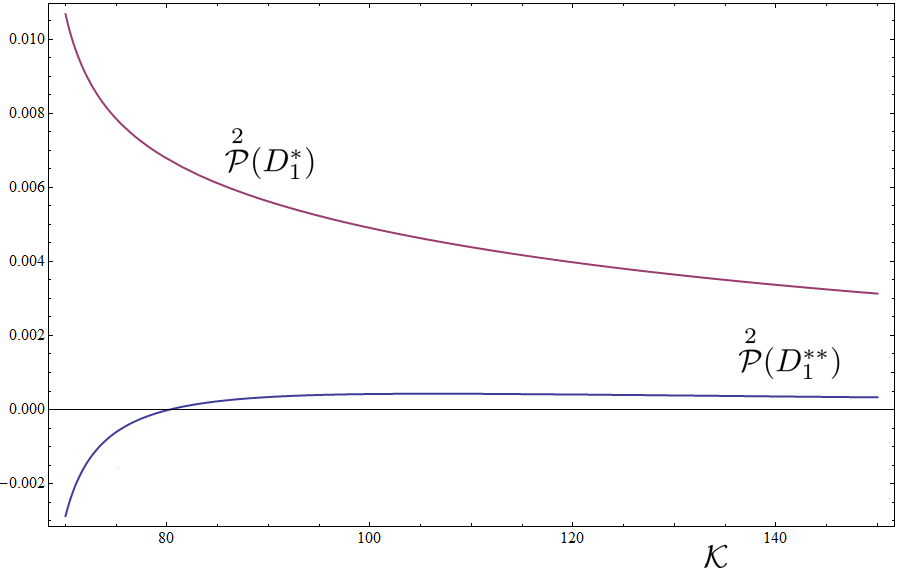}\label{normformD1P}}
\caption{
(a) Values of $\overset{1}{\P}$ for first static branch as a function of parameter $\K$ for a fixed $\Z=1$.
(b) Values of $\overset{2}{\P}$ for the first motile branch ($\Z=1$).} 
\end{center}
\end{figure}

\item \emph{Motile branches} result from pitchfork bifurcations that can be either   supercritical or subcritical with
 $$\P=\overset{0}{\P}+\epsilon^2\overset{2}{\P}/2+o(\epsilon^2). $$ 
The coefficients in this expansion can be written inn the form  
$$
\overset{0}{\P}=-\overset{0}{L}(\overset{0}{L}-1)\quad\text{and}\quad \overset{2}{\P}=\overset{2}{J}-(2\overset{0}{L}-1)\overset{2}{L},
$$
where
$$\overset{2}{J}=\frac{60 \overset{0}{L} \Z-\K (\overset{0}{L}-1) (\overset{0}{L})^3 (\K (\overset{0}{L}-1)-4)}{24 \K (\K (\overset{0}{L}-1)+1)^2}.$$
The length $\overset{0}{L}$ can be found from the system of equations 
$$
\begin{array}{rl}
-\K(\overset{0}{L})^2(\overset{0}{L}-1)&=-\Z\omega^2+(\overset{0}{L})^2\\
\tanh(\omega/2)&=(\omega/2)(1-\Z\omega^2/(\overset{0}{L})^2)
\end{array}
$$
Again, two roots  are  in the interval $[0,1]$: the smaller one belongs to the branch   $D_m^*$ and the larger one to the branch  $D_m^{**}$. 
In Fig.~\ref{normformD1P}, we illustrate the function $\overset{2}{\P}(\K)$  for $m=1$. The bifurcation from the static homogeneous solution with longer length is 
always supercritical as $\overset{2}{\P}(D_1^*)>0$. Instead, the bifurcation from the static homogeneous 
solution with smaller length    can change   from subcritical ($\overset{2}{\mathcal{P}}\leq 0$) 
to supercritical ($\overset{2}{\mathcal{P}}\geq 0$). 
\end{enumerate}

\section{Normal form for the stiff limit.}\label{D}

In the study of  (\ref{eqnstiff}) we closely follow the procedure developed in Section \ref{characteristicequation}. In essence, the results are exactly the same for fixed $\hat{L}=1$ and the product $\mathcal{KP}$ replaced by $\lambda$ with only one homogeneous state ($s(y)=0$, $V=0$ and $s_0=\Phi(r)/r$) and where $\delta s_0$ replace $\delta L$. 
As a result, there is an infinite sequence of bifurcations branching from the now unique homogeneous state. 
We shall only focus on the stable attractor of the problem, 
namely, the homogeneous solution before the $D_1$ bifurcation and the first motile branch after. 

The critical value of the bifurcation parameter $\lambda_c$ corresponding to the case when a homogeneous static solution becomes linearly unstable is given by the formula
 $
\lambda_c=(1+r)^2(1-\mathcal{Z}\omega-c^2).
 $
where $\omega_c$ is a root of the equation
 $
\tanh(\omega_c/2)=\omega_c(1-\mathcal{Z}\omega_c^2)/2
 $
with the smallest absolute value. We then proceed to the next order developing a regular expansions close to the bifurcation point  
$$
\left\{ \begin{array}{c}
s=0+\epsilon\overset{1}{s}+\epsilon^2\overset{2}{s}/2+\epsilon^3\overset{3}{s}/6+o(\epsilon^3)\\
V=0+\epsilon\overset{1}{V}+\epsilon^2\overset{2}{V}/2+\epsilon^3\overset{3}{V}/6+o(\epsilon^3)\\
s_0=\Phi(r)/r+\epsilon\overset{1}{s}_0+\epsilon^2\overset{2}{s}_0/2+\epsilon^3\overset{3}{s}_0/6+o(\epsilon^3),
\end{array} \right.
$$
Similar expansion for the bifurcation has the form
$$
\lambda=\lambda_c+\epsilon\overset{1}{\lambda}+\epsilon^2\overset{2}{\lambda}/2+\epsilon^3\overset{3}{\lambda}/6+o(\epsilon^3).
$$
For the first motile branch one finds that $\overset{1}{\lambda}=0$, indicating a pitchfork bifurcation. 

Below we show that this bifurcation can change  from supercritical to subscritical  depending on the value of the dimensionless  parameter $r$. 
Assuming without loss of generality that $\overset{1}{V}=\overset{2}{V}=1$, we obtain
$$
\overset{2}{\lambda}=\frac{(\omega^2 \Z-1)\left( Ar^2+Br+C\right)}{144 \omega^8 \Z^3 \left(\omega^4 \Z^2-2 \left(\omega^2+6\right) \Z+1\right)}
$$
where
$$
\begin{array}{c}
A=30 \omega^{12} \mathcal{Z}^5-123 \omega^{10} \mathcal{Z}^4+6 \omega^8 (35-164 \mathcal{Z}) \mathcal{Z}^3+2 \omega^6 \mathcal{Z}^2 (1073 \mathcal{Z}-84)+6 \omega^4 \mathcal{Z} \left(1440 \mathcal{Z}^2-155 \mathcal{Z}+8\right)\\
+3 \omega^2 \left(1320 \mathcal{Z}^2-430 \mathcal{Z}+1\right)-9240 \mathcal{Z}+770\\
B=-2 \left(21 \omega^{12} \mathcal{Z}^5-87 \omega^{10} \mathcal{Z}^4+4 \omega^8 (39-173 \mathcal{Z}) \mathcal{Z}^3+2 \omega^6 \mathcal{Z}^2 (707 \mathcal{Z}-66)+3 \omega^4 \mathcal{Z} \left(1440 \mathcal{Z}^2-210 \mathcal{Z}+13\right)\right. \\
\left. +\omega^2 \left(2280 \mathcal{Z}^2-1150 \mathcal{Z}+3\right)-9240 \mathcal{Z}+770\right)\\
C=-6 \omega^{12} \mathcal{Z}^5+21 \omega^{10} \mathcal{Z}^4+6 \omega^8 \mathcal{Z}^3 (28 \mathcal{Z}-5)+2 \omega^6 (12-79 \mathcal{Z}) \mathcal{Z}^2+6 \omega^4 \mathcal{Z} (85 \mathcal{Z}-2)+3 \omega^2 \left(1320 \mathcal{Z}^2-430 \mathcal{Z}+1\right)\\
-9240 \mathcal{Z}+770
\end{array}
$$
\begin{figure}
\centering
\includegraphics[scale=0.2]{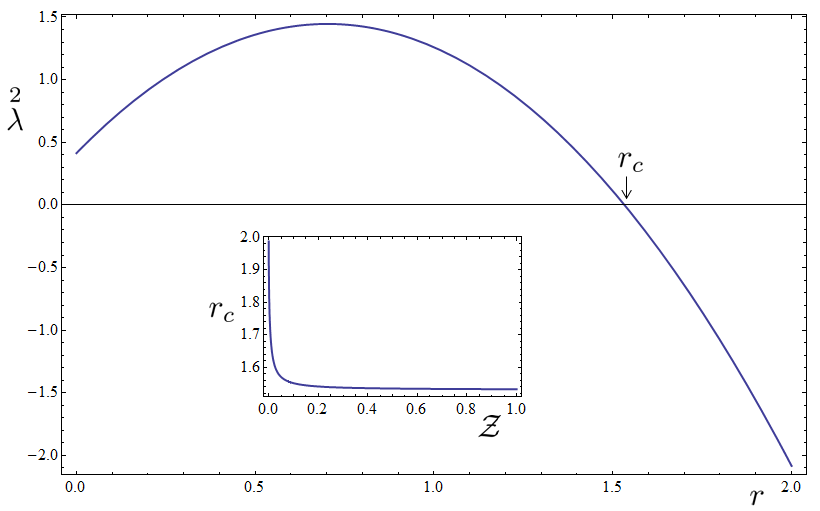}
\caption{\label{nformnonrheology} Parameter $\overset{2}{\lambda}$ characterizing the structure of the  static to motile bifurcation in the case of non linear contraction law. }
\end{figure}
these expressions show that there exists a critical value $r_c$ of the  parameter $r$ such that  the bifurcation is supercritical (i.e. $\overset{2}{\lambda}\leq 0$) for $r\leq r_c$. This regime corresponds to a state where contraction is proportional to concentration of motors.
For $r\geq r_c$, the picthfork bifurcation is subcritical (i.e. $\overset{2}{\lambda}\geq 0$) 
and the regime is characterized by a contraction which saturates into the plateau.
We plot on Fig.~\ref{nformnonrheology} the value of $\overset{2}{\lambda}$ as a function of $r$ 
for a fixed $\Z=1$ and in inset the value $r_c(\Z)$. When $\Z\rightarrow 0$, $r_c \rightarrow 2$ 
and when $\Z\rightarrow \infty$,  $r_c \rightarrow (7+\sqrt{69})/10$.


\end{document}